\documentclass[prd,reprint,nofootinbib,superscriptaddress,preprintnumbers,showpacs]{revtex4-1}
\usepackage{amsfonts}
\usepackage{amssymb}
\usepackage{amsmath}
\usepackage{graphicx,color}
\usepackage{dcolumn}
\usepackage{epsfig}
\usepackage{bm}
\usepackage{bbm}
\usepackage{ulem}
\usepackage{hyperref}
\usepackage{lineno}
\usepackage{multirow}

\usepackage{color}

\def\gtap{\ \raise.3ex\hbox{$>$\kern-.75em\lower1ex\hbox{$\sim$}}\ }
\def\ltap{\ \raise.3ex\hbox{$<$\kern-.75em\lower1ex\hbox{$\sim$}}\ }
\begin{document}

\title{Three-body
unitary coupled-channel approach to radiative $J/\psi$ decays
and $\eta(1405/1475)$
}
\author{S.X. Nakamura}
\email{satoshi@sdu.edu.cn}
\affiliation{
Institute of Frontier and Interdisciplinary Science, Shandong
University, Qingdao, Shandong 266237, China
}
\affiliation{
University of Science and Technology of China, Hefei 230026, 
China
}
\affiliation{
State Key Laboratory of Particle Detection and Electronics (IHEP-USTC), Hefei 230036, China}
\author{Q. Huang}
\affiliation{
Department of Physics, Nanjing Normal University, Nanjing, Jiangsu 210097, China
}
\author{J.-J. Wu}
\email{wujiajun@ucas.ac.cn}
\affiliation{
School of Physical Sciences, University of Chinese Academy of Sciences (UCAS), Beijing 100049, China
}
\author{H.P. Peng}
\author{Y. Zhang}
\author{Y.C. Zhu}
\affiliation{
University of Science and Technology of China, Hefei 230026, 
China
}
\affiliation{
State Key Laboratory of Particle Detection and Electronics (IHEP-USTC), Hefei 230036, China}

\begin{abstract}
Recent BESIII data on radiative $J/\psi$ decays from
$\sim 10^{10}$ $J/\psi$ samples should significantly advance our
understanding of the controversial nature of $\eta(1405/1475)$.
This motivates us to
develop a three-body unitary coupled-channel model for radiative
$J/\psi$ decays to three-meson final states of any partial wave ($J^{PC}$).
Basic building blocks of 
the model are bare  resonance states such as $\eta(1405/1475)$ and $f_1(1420)$,
and $\pi K$, $K\bar{K}$, and
$\pi\eta$ two-body interactions that generate resonances such as
$K^*(892)$, $K^*_0(700)$, and $a_0(980)$.
This model reasonably fits
$K_SK_S\pi^0$ Dalitz plot pseudo data 
generated from 
the BESIII's $J^{PC}=0^{-+}$ amplitude for $J/\psi\to\gamma K_SK_S\pi^0$.
The experimental branching ratios of 
$\eta(1405/1475)\to\eta\pi\pi$ and 
$\eta(1405/1475)\to\gamma\rho$ 
relative to that of $\eta(1405/1475)\to K\bar{K}\pi$ 
are simultaneously fitted. 
Our $0^{-+}$ amplitude is analytically continued to find three poles,
two of which correspond to $\eta(1405)$ on different Riemann sheets of the
$K^*\bar{K}$ channel, and the third one for $\eta(1475)$.
This is the first pole determination of $\eta(1405/1475)$ and, furthermore, 
the first-ever pole determination from analyzing experimental Dalitz plot distributions
with a manifestly three-body unitary coupled-channel framework.
Process-dependent 
$\eta\pi\pi$, $\gamma\pi^+\pi^-$, and $\pi\pi\pi$ lineshapes of 
$J/\psi\to\gamma(0^{-+})\to \gamma(\eta\pi\pi)$, $\gamma(\gamma\rho)$, and
$\gamma(\pi\pi\pi)$ are predicted, and are 
in reasonable agreement with data. 
A triangle singularity is shown to play 
a crucial role to cause the large isospin violation of
$J/\psi\to\gamma(\pi\pi\pi)$.
\end{abstract}

\maketitle

\section{introduction}
\label{sec:intro}

Since the first
observation in 1967~\cite{Baillon:1967zz},
the light isoscalar pseudoscalar states in 1.4--1.5~GeV region, named
$\eta(1405/1475)$, has invited lots of debates about its peculiar features
in experimental data and about various theoretical interpretations.
Two major questions on $\eta(1405/1475)$, which are still open, are: (i) Are there one or
two $\eta$ excited states in this energy region ?; 
(ii) How is the internal structure of the excited state(s) like ?
What makes $\eta(1405/1475)$ difficult to understand is that
$\eta(1405/1475)$ could include various components such as
a quark-antiquark pair, various hadronic
coupled-channels, and a glueball, reflecting the complex nature
of the Quantum Chromodynamics (QCD) in the low energy regime.
Also, the mixing between $(u\bar{u}+d\bar{d})/\sqrt{2}$ and $s\bar{s}$
is significant only in the isoscalar pseudoscalar sector.
Thus, understanding $\eta(1405/1475)$ seems particularly important to deepen our understanding of QCD.

$\eta(1405/1475)$ has been seen in various processes.
However, the $\eta(1405/1475)$ lineshapes appear rather process different and thus have been explained
with a single or two different states.
For example, a single peak appears in $\eta\pi\pi$ final states from $p\bar{p}$ annihilation~\cite{amsler}, radiative $J/\psi$ decays~\cite{mark3_jpsi-gamma-eta-pipi,bes_jpsi-gamma-eta-pipi,dm2_jpsi-gamma-eta-pipi}, and $J/\psi\to\omega(\eta\pi\pi)$~\cite{bes3_jpsi-omega-eta-pipi} at somewhat process-dependent peak positions.
A single peak is also found in $K\bar{K}\pi$ and $\eta\pi\pi$ final states from $\gamma\gamma$
collisions~\cite{L3}, and $\gamma\rho^0$ final states from radiative $J/\psi$ decays~\cite{bes2_rhog,mark3_rhog,dm2_jpsi-gamma-eta-pipi} and from
$p\bar{p}$ annihilation~\cite{amsler}.
On the other hand, structures seemingly due to two overlapping resonances are seen in $K\bar{K}\pi$ invariant mass distributions in $\pi^-p$ scattering~\cite{e852,e769}, $p\bar{p}$ annihilations~\cite{obelix2002}, and radiative $J/\psi$ decays~\cite{mark3_1990,dm2_1992}.

The conventional quark model expects radially excited $\eta$ and $\eta'$ states in this energy region, and they correspond to $\eta(1295)$ and (one of) $\eta(1405/1475)$ states, respectively~\cite{pdg,barnes1997}. 
If $\eta(1405/1475)$ includes two states, what is its nature ?
A proposal was made to interpret $\eta(1405)$ as a glueball~\cite{faddeev2004}.
However, the isoscalar pseudoscalar glueball from lattice QCD (LQCD) predictions turned out to be significantly heavier~\cite{bali1993,morningstar1999,chen2006,richards2010,chen2111}.
Meanwhile, a LQCD prediction from Ref.~\cite{dudek2013} indicated only
two states in this region. 
However, the authors did not identify them with $\eta(1295)$ and
$\eta(1405/1475)$ since the experimental situation is unclear.
Thus, although the two-state solution for $\eta(1405/1475)$ is not accommodated in the quark
model, it is not forbidden by any strong theoretical arguments.

Another peculiar property of $\eta(1405/1475)$ is its anomalously large
isospin violation in $\eta(1405/1475)\to\pi\pi\pi$ decays, as found 
in radiative $J/\psi$ decays
by the BESIII collaboration in 2012~\cite{bes3-3pi}.
The BESIII found that the decays mostly proceed as $\eta(1405/1475)\to f_0(980)\pi\to \pi\pi\pi$,
and that the rate is significantly larger than that expected from
$\eta(1405/1475)\to a_0(980)\pi$ followed by the $a_0(980)$-$f_0(980)$ mixing.
It was also found that the $f_0(980)$ width in the $\pi\pi$ invariant
mass distribution is significantly narrower ($\sim 10$~MeV) than those
seen in other processes ($\sim 50$~MeV)~\cite{pdg}.
A theoretical explanation for these experimental findings was made 
in Refs.~\cite{wu2012,wu2013,Aceti:2012dj,du2019}.
First the authors pointed out that a $K^*\bar{K}K$ triangle loop from a $\eta(1405/1475)$
decay can hit an on-shell kinematics, causing a triangle singularity
(TS) that can significantly enhance the amplitude.
At the same time, this triangle loop causes the isospin violation
due to the mass difference between $K^\pm$ and $K^0$ 
in the $K^*K^-K^+$ and $K^*\bar{K}^0K^0$ triangle loops.
This mechanism can naturally explain the large isospin violation 
 without any additional assumptions. 

The discovery of 
the potentially important TS effects in the $\eta(1405/1475)$ decays
encouraged theorists to describe all $\eta(1405/1475)$-related
data, including process-dependent lineshapes, with one $\eta(1405/1475)$ state,
based on the ``Occam's Razor'' principle~\cite{wu2012,wu2013,Aceti:2012dj,du2019}.
Indeed, it was shown that the TS mechanisms can shift 
the resonant peak position somewhat, depending on $K\bar{K}\pi$, $\eta\pi\pi$,
and $\pi\pi\pi$ final states.
However, the experimental data of $K\bar{K}\pi$
and $\eta\pi\pi$ were rather limited at this time, 
and these theoretical results were not sufficiently tested.
Also, it has not been possible to discriminate one- and two-state solutions of 
$\eta(1405/1475)$.

A significant advancement has been made recently by a BESIII analysis of
$J/\psi\to\gamma(K_SK_S\pi^0)$ data
from the high-statistics $\sim 10^{10} J/\psi$ decay samples~\cite{bes3_mc}.
They fitted the data with $J^{PC}=0^{-+}$, $1^{++}$, and $2^{++}$ partial wave amplitudes,
and identified two $\eta(1405/1475)$ states in the $0^{-+}$ amplitude with a high statistical significance.

There are however theoretical issues in the BESIII analysis since they described the $\eta(1405/1475)$ states with Breit-Wigner (BW) amplitudes.
The BW amplitude is known to be unsuitable in cases when a resonance is close to its decay channel threshold and/or when more than one resonances are overlapping~\cite{3pi-2}.
This difficulty arises since the BW amplitude does not consider the unitary.
In the present case, $\eta(1405)$ is close to the $K^*\bar{K}$ threshold, and $\eta(1405)$ and $\eta(1475)$ are overlapping.
Furthermore, while 
coupling parameters in the BW formalism
implicitly absorb loop contributions, they cannot simulate non-smooth behavior
such as TS.
Thus, it is highly desirable to develop an appropriate approach where
the data are fitted with a unitary coupled-channel $J/\psi$ decay amplitude, and 
$\eta(1405/1475)$ poles are searched by analytically continuing the
amplitude.
The $\eta(1405/1475)$ exists in a complicated coupled-channel system consisting of quasi two-body channels such as $K^*\bar{K}$ and $a_0\pi$ and three-body channels such as $K\bar{K}\pi$ and $\pi\pi\eta$.
The unitary coupled-channel approach seems the only possible option to describe such a system.
Also in this approach, we automatically take account of the TS effects
that are expected to play an important role,
and thus
taking over the sound physics in the previous models of
Refs.~\cite{wu2012,wu2013,Aceti:2012dj,du2019}.


In this work~\footnote{A part of the results has been published in Ref.~\cite{letter}.},
we develop a three-body unitary coupled-channel model for radiative
$J/\psi$ decays to three-meson final states of any $J^{PC}$.
Then we use the model to fit $K_SK_S\pi^0$ Dalitz plot pseudodata generated from 
the BESIII's $0^{-+}$ amplitude for $J/\psi\to\gamma(K_SK_S\pi^0)$~\cite{bes3_mc}.
At the same time, the branching fractions of other final states such as 
$\eta\pi^+\pi^-$ and $\rho^0\gamma$ relative to that of $K\bar{K}\pi$ are also fitted.
Based on the obtained model, 
we examine the pole structure of $\eta(1405/1475)$ in the complex
energy plane to see if $\eta(1405/1475)$ is 
one or two state(s).
We also use the model to predict $\eta(1405/1475)\to \eta\pi\pi$, $\gamma\pi\pi$, and
$\pi\pi\pi$ lineshapes and branchings.
By examining the $\eta(1405/1475)$ decay mechanisms for different final
states, we identify dominant mechanisms
and address major issues regarding $\eta(1405/1475)$ how the
process dependent lineshapes and large isospin violations come about.

Precise Dalitz plot data are a great target for a three-body unitary model.
Single-channel three-body unitary frameworks based on  the Khuri-Treiman equations have been used extensively to analyze Dalitz data in elastic kinematical regions: e.g., Refs.~\cite{KT1,KT2} for $\omega/\phi\to \pi\pi\pi$.
However, Dalitz-plot analyses covering inelastic kinematical regions with coupled-channel three-body unitary frameworks are very limited: e.g., Ref.~\cite{d-decay} for $D^+\to K^-\pi^+\pi^+$ and the present analysis.
Since more and more precise Dalitz data are expected from the contemporary experimental facilities, the importance of the three-body unitary coupled-channel analysis will be increasing. 
Thus, related theoretical developments have been made recently~\cite{3pi,three-body1,three-body2}.

Three-body unitary analysis like the present work involves pole extractions.
There are literatures~\cite{gloeckle,flinders,julich-a1} that discuss the pole extraction from 
three-body unitarity amplitudes.
Practically, however, such a pole extraction from experimental three-body distributions had not been done until recently. 
The first case was made in Refs.~\cite{a1-jpac,a1-gwu} where a $\rho\pi$ single-channel model was used to analyze $m_{\pi^+\pi^-\pi^-}$ lineshape data for $\tau^-\to\pi^+\pi^-\pi^-\nu_\tau$, extracting an $a_1(1260)$ pole.
Reference~\cite{a1-gwu} (\cite{a1-jpac}) treated the three-body unitarity rigorously (partially).
The two analyses highlighted the importance of the full three-body unitarity in the pole extraction since an additional spurious pole existed in Ref.~\cite{a1-jpac}.
Our present analysis treats the three-body unitarity as rigorously as in Ref.~\cite{a1-gwu}.
Furthermore, we improve the pole extraction method of Ref.~\cite{a1-gwu} since we consider relevant coupled-channels and fit Dalitz plot distributions rather than the projected invariant mass distributions.

The organization of this paper is as follows.
In Sec.~\ref{sec:model}, we present formulas for the radiative $J/\psi$ decay amplitude based on the three-body unitary coupled-channel model and the partial decay width.
In Sec.~\ref{sec:analysis}, we analyze Dalitz plot pseudodata from the BESIII $0^{-+}$ amplitude for $J/\psi\to\gamma(K_SK_S\pi^0)$.
The quality of the fits is shown, and the $\eta(1405/1475)$ poles are extracted.
In Sec.~\ref{sec:prediction}, we predict the lineshapes of $\eta\pi\pi$, $\gamma\pi^+\pi^-$, $\pi\pi\pi$ final states from the radiative $J/\psi$ decays.
The branching fractions for the  $\pi\pi\pi$ final states are also predicted.
Finally, in Sec.~\ref{sec:summary}, we summarize the paper, and discuss the
future prospects.

\section{model}
\label{sec:model}

\subsection{Radiative $J/\psi$ decay amplitudes within three-body unitary
  coupled-channel approach}\label{sec:model-1}

\begin{figure}
\begin{center}
\includegraphics[width=.5\textwidth]{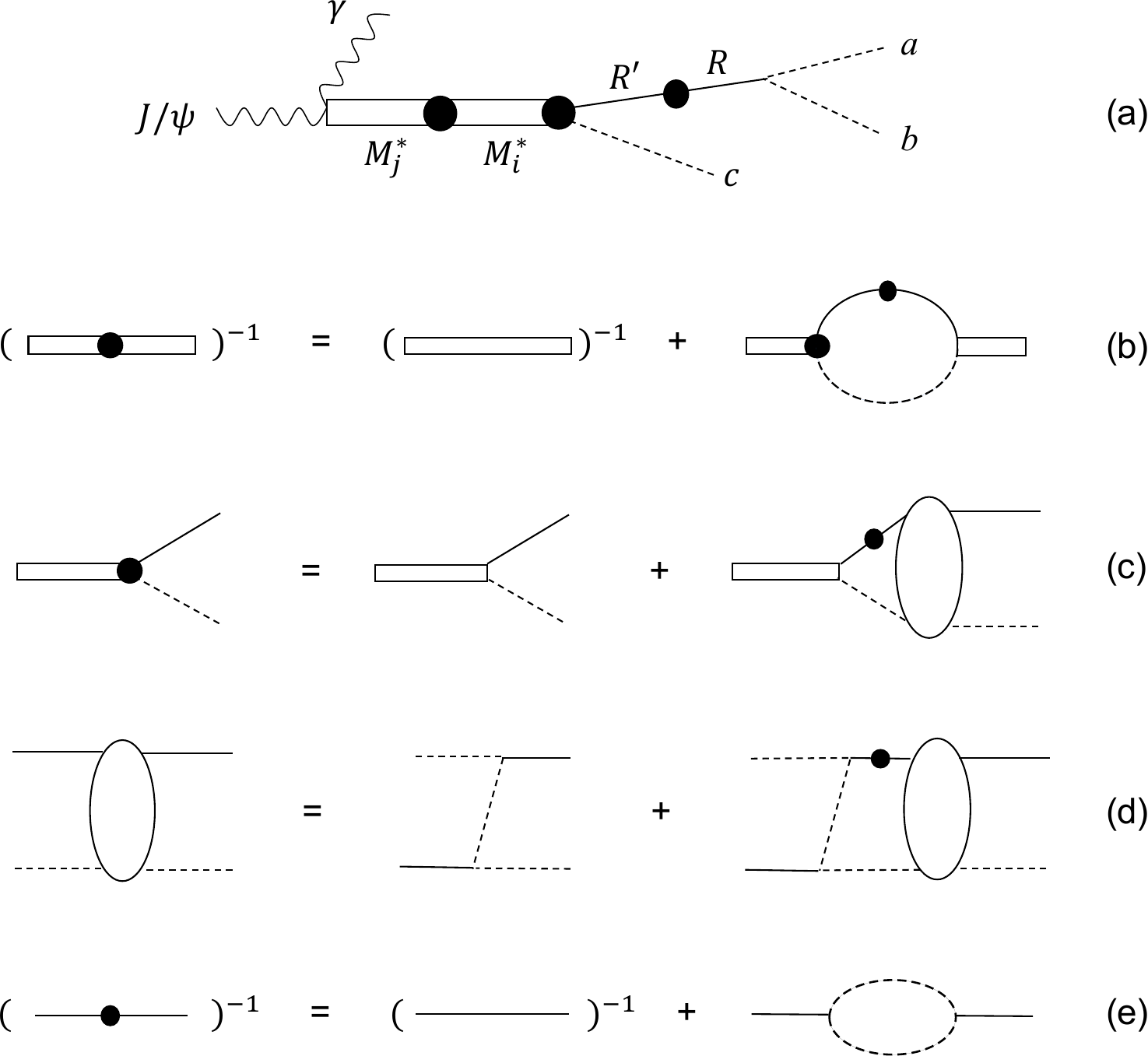}
\end{center}
 \caption{
(a) Diagrammatic representation for 
radiative $J/\psi$ decay amplitude of Eq.~(\ref{eq:amp_a}).
The dashed lines represent pseudoscalar mesons 
while the solid lines are bare two-meson resonances $R$.
 The double lines with $M^*_{i(j)}$ represent bare states for $M^*$ such
 as $\eta(1405/1475)$.
(b) The dressed $M^*$ propagator: the first [second] diagram on
 the r.h.s. is a bare $M^*$ propagator [self energy].
(c) The dressed $M^*$ decay vertex: the first [second] diagram
is a bare vertex [rescattering term].
The ellipse stands for the scattering amplitude $X$ 
(d) Lippmann-Schwinger-like equation for the amplitude $X$.
(e) The dressed $R$ propagator: the first [second] diagram
is a bare $R$ propagator [self energy].
 }
\label{fig:diag}
\end{figure}

In constructing 
our three-body unitary coupled-channel model,
we basically follow the formulation presented in Refs.~\cite{3pi,d-decay}.
However, there is one noteworthy difference.
While we specified a particle with its isospin state in Refs.~\cite{3pi,d-decay},
we now use its charge state.
This is an important extension of the model to describe
isospin-violating processes. 
In what follows, we sketch 
our model,
putting an emphasis on the differences from Refs.~\cite{3pi,d-decay}.

A radiative $J/\psi$ decay mechanism within our model is 
diagrammatically represented by Fig.~\ref{fig:diag}(a).
First, $J/\psi$ radiatively couples,
via a vertex $\Gamma_{\gamma M^*_j,J/\psi}$,
 to a bare excited state 
($M^*$) such as $\eta(1405/1475)$ of $J^{PC}=0^{-+}$ and
$f_1(1420)$ of $J^{PC}=1^{++}$;
we consider $M^*$ with $I=0$ ($I$: isospin) in this work. 
Second, the bare $M^*$ nonperturbatively couples with quasi two-body $Rc$ 
and three-body $abc$ states to form 
a dressed $M^*$ propagator $\bar{G}_{M^*}$ [Fig.~\ref{fig:diag}(b)]
that includes $M^*$ resonance pole(s).
Here, $abc$
are pseudoscalar mesons ($\pi$, $K$, $\eta$), 
and $R$ is a bare two-meson resonance such as $K^*$, $a_0(980)$, and
$f_0(980)$.
Particles $R$ and $ab$ also 
nonperturbatively couple through a vertex $\Gamma_{ab,R}$,
forming a dressed $R$ propagator $\tau_{R,R'}$
[Fig.~\ref{fig:diag}(e)] that includes 
$R$ resonance pole(s).
Third, $M^*$ decays to a final $abc$ via a dressed $M^*_i\to Rc$ decay
vertex $\bar{\Gamma}_{cR,M^*_i}$ [Fig.~\ref{fig:diag}(c)] that includes
nonperturbative final state interactions. 
The amplitude formula for the above radiative $J/\psi$ decay process
is given by~\footnote{
We denote a particle $x$'s mass, momentum, energy, polarization, spin
and its $z$-component in the $abc$ center-of-mass frame
by $m_x$, $\bm{p}_x$, $E_x$, $\bm{\epsilon}_x$, $s_x$, and $s_x^z$,
respectively;
$E_x=\sqrt{m_x^2+p_x^2}$ with $p_x=|\bm{p}_x|$.
The mass values for pseudoscalar mesons ($\pi$, $K$, $\eta$)
are taken from Ref.~\cite{pdg}.
Symbols with tilde such as $\tilde{\bm{p}}_x$ indicate 
quantities in the $J/\psi$-at-rest frame.
}
\begin{eqnarray}
 A_{\gamma abc,J/\psi} &=& \sum_{J^{PC}} A^{J^{PC}}_{\gamma abc,J/\psi} \ ,
\label{eq:amp_a_sum}
\end{eqnarray}
with
\begin{eqnarray}
A^{J^{PC}}_{\gamma abc,J/\psi} &=& 
\sum^{\rm cyclic}_{abc}
\sum_{RR's_R^z}
\sum_{ijs_{M^*}^z}
\Gamma_{ab,R}\,
\tau_{R,R'}(p_c,E-E_{c})\,
\nonumber\\
&&\times \bar{\Gamma}_{cR',M^*_i}(\bm{p}_c, E)\,
[\bar{G}_{M^*}(E)]_{ij}\,
\Gamma_{\gamma M^*_j,J/\psi}\, ,
\label{eq:amp_a}
\end{eqnarray}
where cyclic permutations $(abc), (cab), (bca)$ are indicated by 
$\sum^{\text{cyclic}}_{abc}$;
the indices $i$ and $j$ specify one of bare $M^*$ states belonging to
the same $J^{PC}$;
$E$ denotes the $abc$ total energy
in the $abc$ center-of-mass (CM) frame.
Below, 
we give a more detailed expression for each of the components in the amplitude.

The $J/\psi\to \gamma M^*_j$ vertex is given in a general form as 
\begin{eqnarray}
\Gamma_{\gamma M^*_j,J/\psi}
&=& 
\sum_{\ell\tilde{\ell}^z s}
{g^{\ell s}_{J/\psi M^*_j\gamma}\,
(s_{M^*} \tilde{s}^z_{M^*} 1 \tilde{s}^z_\gamma |s \tilde{s}^z_{M^*}+\tilde{s}^z_\gamma)
\over \sqrt{8 \tilde{E}_\gamma\, m_{J/\psi}\, m_{M^*_j}}}
\nonumber\\ 
&&\times (\ell \tilde{\ell}^z s \tilde{s}^z_{M^*}+\tilde{s}^z_\gamma |1 \tilde{s}^z_\psi)
Y_{\ell \tilde{\ell}^z}(\hat{\tilde{p}}_\gamma) \tilde{p}^\ell_\gamma \ ,
\label{eq:invamp2p}
\end{eqnarray}
where $g^{\ell s}_{J/\psi M^*_j\gamma}$ is a coupling constant and 
$m_{M^*_j}$ being a bare $M^*_j$ mass;
$Y_{\ell m}(\hat{q})$ denotes the spherical harmonics with 
$\hat{q}\equiv \bm{q}/|\bm{q}|$, and 
$\sum_\ell$ is restricted by the parity-conservation.
When $M^*$ belongs to $J^{PC}=0^{-+}$, Eq.~(\ref{eq:invamp2p}) reduces
to (up to a constant overall factor)
\begin{eqnarray}
\Gamma_{\gamma \eta^*_j,J/\psi}
&=& 
{g_{J/\psi \eta^*_j\gamma}\,
(\tilde{\bm{\epsilon}}_{J/\psi}\times \tilde{\bm{\epsilon}}_{\gamma})\cdot \tilde{\bm{p}}_\gamma
\over \sqrt{8 \tilde{E}_\gamma\, m_{J/\psi}\, m_{\eta^*_j}}}
\ .
\label{eq:invamp2}
\end{eqnarray}
In our numerical analysis from Sec.~\ref{sec:analysis}, 
we use the coupling $g_{J/\psi \eta^*_j\gamma}$ defined in this reduced form.
The $R\to ab$ vertex is given by
\begin{eqnarray}
\Gamma_{ab,R}&=&
(t_a t^z_a t_b t^z_b |t_R t^z_R)
\sum_{LL^zSS^z}  
(s_a s^z_a s_b s^z_b | S S^z) 
\nonumber\\
&&\times
(L L^z S S^z |s_R s^z_R )
Y_{LL^z}(\hat{p}^*_a) 
\nonumber\\
&&\times
\sqrt{m_R E_a(p_a^*) E_b(p_a^*)
\over E_R(p_c) E_a(p_a) E_b(p_b)}
f^{LS}_{ab,R}(p_a^*) ,
\label{eq:pipi-vertex0-general}
\end{eqnarray}
where the parentheses are Clebsch-Gordan coefficients, and
$t_x$ and $t_x^z$
are the isospin of a particle $x$ and its $z$-component, respectively;
$\bm{p}_a^*$ denotes a particle $a$'s momentum 
in the $ab$ CM frame.
Since particles $a$ and $b$ are pseudoscalar in this paper,
the total spin is $S=0$ and the orbital angular momentum is 
$L=s_R$.
Thus we
simplify the above notation for the $R\to ab$ vertex as
\begin{eqnarray}
\Gamma_{ab,R}&=&
(t_a t^z_a t_b t^z_b |t_R t^z_R)
Y_{s_R,s^z_R}(\hat{p}^*_a) 
\nonumber\\
&&\times
\sqrt{m_R E_a(p_a^*) E_b(p_a^*)
\over E_R(p_c) E_a(p_a) E_b(p_b)}
f_{ab,R}(p_a^*) ,
\label{eq:pipi-vertex0}
\end{eqnarray}
with a vertex function
\begin{eqnarray}
f_{ab,R}(q)=
g_{ab,R} {
 (1+q^2/c_{ab,R}^2)^{-2-{L\over 2}}
\over \sqrt{m_R E_a(q) E_b(q)}} {q^L\over m_\pi^{L-1}} ,
\label{eq:pipi-vertex}
\end{eqnarray}
and use this notation hereafter. 
The coupling 
$g_{ab,R}$ and cutoff $c_{ab,R}$ 
in Eq.~(\ref{eq:pipi-vertex}), 
and the bare mass $m_R$ in 
Eq.~(\ref{eq:green-Rc})
are determined by analyzing
$L$-wave $ab$ scattering data
as detailed in Appendix~\ref{app1} where the parameter values are presented.

The dressed $R$ propagator [Fig.~\ref{fig:diag}(e)] is given by
\begin{eqnarray}
[ \tau^{-1}(p,E) ]_{R,R'}
&=& 
[ E - E_{R}(p) ] \delta_{R,R'} 
- \Sigma_{R,R'}\left(p, E\right) ,
\label{eq:green-Rc}
\end{eqnarray}
with the $R$ self-energy 
\begin{eqnarray}
\Sigma_{R,R'}(p,E) &=& 
\sum_{ab} 
(t_a t^z_a t_b t^z_b |t_R, t^z_a + t^z_b)
(t_a t^z_a t_b t^z_b |t_{R'}, t^z_a + t^z_b)
\nonumber\\
&&\times
\sqrt{\frac{m_{R}m_{R'}}{E_{R}(p)E_{R'}(p)}}
\int q^2 dq
{M_{ab}(q)\over \sqrt{M^2_{ab}(q) + p^2}}
\nonumber \\
&&\times
\frac{ {\cal B}_{ab}\, f_{R, ab}(q) f_{ab,R'}(q)}
{E - \sqrt{M^2_{ab}(q) + p^2} + i\epsilon} ,
\label{eq:RR-self}
\end{eqnarray}
with $M_{ab}(q)=E_{a}(q)+E_{b}(q)$ and $m_R$ being a bare mass of $R$;
$R$ and $R'$ in Eq.~(\ref{eq:RR-self}) have the same spin state
($s_R=s_{R'}$).
Due to the Bose symmetry, we have a factor ${\cal B}_{ab}$:
${\cal B}_{ab}=1/2$ for identical particles $a$ and $b$;
${\cal B}_{ab}=1$ otherwise.
In Eq.~(\ref{eq:RR-self}), 
the $a_0$-$f_0$ mixing occurs ($\Sigma_{a_0,f_0}\ne 0$)
due to the mass difference between 
$ab=K^+K^-$ and $K^0\bar{K}^0$ states.
The dressed $R$ propagators include $R$ resonance poles as summarized in
Tables~\ref{tab:pole-pipi}--\ref{tab:pole-pieta} in Appendix~\ref{app1}.

The dressed $M^*_i\to Rc$ decay vertex [Fig.~\ref{fig:diag}(c)] is
given by 
\begin{eqnarray}
\bar{\Gamma}_{cR,M^*_i} (\bm{p}_c, E)
&=&
\sum_{ l, l^z}
(l l^z s_R s^z_R | s_{M^*} s^z_{M^*})
(t_R t^z_R t_c t^z_c |t_{M^*} t^z_{M^*})\nonumber \\
&&\times Y_{l,l^z}(-\hat p_c) \bar F_{(cR)_l,M^*_i}(p_c, E),
\label{eq:dressed_mstar}
\end{eqnarray}
where $l$ is the relative orbital angular momentum between $R$ and $c$.
The dressed $M^*_i\to Rc$ vertex function is 
\begin{eqnarray}
\label{eq:dressed-g}
\bar F_{(cR)_l ,M^*_i}(p_c ,E)&=&
F_{(cR)_l , M^*_i}(p_c)
+ \sum_{c'R'R''l'} \int q^2 dq 
\nonumber \\
&& \times \,
 X^{J^{PC}}_{(cR)_{l},(c'R'')_{l'}}(p_c,q;E) 
\nonumber \\
&&\times \,
\tau_{R'',R'}(q,E-E_{c'}) F_{(c'R')_{l'},M^*_i}(q) ,
\nonumber\\
\end{eqnarray}
where the first and second terms are direct decay and rescattering
mechanisms, respectively.
Common isobar models do not have the second term.
We use a bare vertex function including a dipole form factor as
\begin{eqnarray}
F_{(cR)_{l},M^*_i}(q)\! &=& 
C^{M^*_i}_{(c R)_l}\!
\frac{ [1+q^2/(\Lambda^{M^*_i}_{(c R)_l})^2]^{-2-{l\over 2}} 
}
{\sqrt{8 E_c(q) E_R(q)\, m_{M^*_i}}} 
{q^l\over m_\pi^{l-1}},
\label{eq:bare_mstar}
\end{eqnarray}
where $C^{M^*_i}_{(c R)_l}$ and 
$\Lambda^{M^*_i}_{(c R)_l}$ are 
coupling and cutoff parameters, respectively.
We also have introduced $J^{PC}$ partial wave amplitudes for
$cR\to c'R'$ scatterings, $X^{J^{PC}}_{(cR)_{l},(c'R')_{l'}}$,
that is obtained by solving the scattering
equation [Fig.~\ref{fig:diag}(d)]:
\begin{eqnarray}
&&X^{J^{PC}}_{(c'R')_{l'},(cR)_{l}} (p',p; E)
\nonumber\\
&&=
V^{J^{PC}}_{(c'R')_{l'},(cR)_{l}} (p',p; E)
\nonumber\\
&&
+ \sum_{c'',R'',R''',l''}
\int q^2dq\, V^{J^{PC}}_{(c'R')_{l'},(c''R''')_{l''}}(p',q;E) 
\nonumber\\
&&
\qquad \times \tau_{R''',R''}(q,E-E_{c''}) 
X^{J^{PC}}_{(c''R'')_{l''},(cR)_{l}}(q,p;E) ,
\label{eq:pw-tcr}
\end{eqnarray}
with
\begin{eqnarray}
V^{J^{PC}}_{(c'R')_{l'},(cR)_{l}} (p',p; E)
&=&Z^{\bar{c},J^{PC}}_{(c'R')_{l'},(cR)_{l}} (p',p; E)
\nonumber\\
&&+v^{{\rm HLS},J^{PC}}_{(c'R')_{l'},(cR)_{l}} (p',p) .
\label{eq:vvv}
\end{eqnarray}
The driving term $Z^{\bar{c},J^{PC}}_{(c'R')_{l'},(cR)_{l}}$, what we call the $Z$-diagram, is
diagrammatically expressed in 
the first term of the r.h.s. of Fig.~\ref{fig:diag}(d);
$\bar{c}$ indicates an exchanged particle.
Explicit formulas for the partial-wave-expanded
$Z$-diagram can be found in Appendix~C of Ref.~\cite{3pi}.
One important difference from Ref.~\cite{3pi} is that 
we here do not project the $Z$ diagrams onto a definite total isospin
state. As a result, 
an isospin-violating $K^*\bar{K}\to f_0\pi$ process is caused by a
$Z$-diagram and $m_{K^\pm}\ne m_{K^0}$, leading to $\eta^*\to \pi\pi\pi$.

The second term in the r.h.s. of Eq.~(\ref{eq:vvv})
is a vector-meson exchange mechanism based on the hidden local symmetry model~\cite{hls}.
In the present case, this mechanism works for $K^*\bar{K}\leftrightarrow K^*\bar{K}, \bar{K}^*K$
interactions. 
Formulas are presented in Appendix~A of Ref.~\cite{d-decay}, but here we 
use a different form factor of
$(1+p^2/\Lambda^2)^{-2}(1+p'^2/\Lambda^2)^{-2}$ with $\Lambda=1$~GeV,
rather than Eq.~(A15) of Ref.~\cite{d-decay}.

The dressed $M^*$ propagator [Fig.~\ref{fig:diag}(b)]
 is given by 
\begin{eqnarray}
\left[\bar{G}_{M^*}^{-1}(E)\right]_{ij} = (E- m_{M^*_i})\delta_{ij} - \left[\Sigma_{M^*}(E)\right]_{ij}\,,
\label{eq:mstar-g1}
\end{eqnarray}
where the $M^*$ self energy in 
the second term is given by 
\begin{eqnarray}
[\Sigma_{M^*}(E)]_{ij} &=& 
{\cal B}_{Rc}
\sum_{cRR'l} 
\int q^2 dq\,
F_{(cR)_l ,M^*_i}(q)
\nonumber\\
&&\times 
\tau_{R,R'}(q,E-E_c(q)) 
\bar F_{(cR')_l ,M^*_j}(q ,E)  .
\label{eq:mstar-sigma}
\end{eqnarray}
The above formulas show that 
the dressed $M^*$ propagator ($M^*$ pole structure)
and the dressed $M^*_i\to Rc$ form factor ($M^*$ decay mechanism)
are explicitly related by the common dynamics.
This is a consequence of the three-body unitarity.

In the above formulas, we assumed that 
two-body $ab\to a'b'$ interactions occur via bare $R$-excitations,
$ab\to R\to a'b'$.
We can straightforwardly extend the formulas if
two-body interactions are from 
bare $R$-excitations and separable contact interactions, 
as detailed in Ref.~\cite{d-decay}.
Also, the above formulas are valid when $c$ is 
a pseudoscalar meson, and need to be slightly modified
for $Rc=\rho\rho$ channel.
We consider the spectator $\rho$ width in
the first term of the r.h.s. of Eq.~(\ref{eq:green-Rc})
by $E - E_{R}(p) \to E - E_{R}(p) + i \Gamma_\rho/2$;
$\Gamma_\rho=150$~MeV and is constant.
Also, the label in the bare form factor of Eq.~(\ref{eq:bare_mstar}) is
extended to include the total spin of $\rho\rho$ ($s_{\rho\rho}$) as
$(c R)_l\to (c R)_{ls_{\rho\rho}}$.

For describing $J/\psi\to\gamma M^*\to\gamma(\gamma\pi^+\pi^-)$,
we assume the vector-meson dominance mechanism
where the $\rho\rho$ channel from 
$M^*_i\to\rho\rho$ or coupled-channel dynamics
is followed by $\rho\to\gamma$ and $\rho\to\pi^+\pi^-$.
The photon-$\rho$ direct coupling is from 
the vector-meson dominance model.
This mechanism can be implemented in the decay amplitude formula of
Eq.~(\ref{eq:amp_a}) by multiplying $2 e/g_\rho$;
each of two $\rho$ can couple to the photon, giving 
a factor of 2; 
$e^2/4\pi\simeq 1/137$ and $g_\rho^2/4\pi=2.2$.
There are some experimental indications for 
$\eta(1405/1475)\to\rho\rho\to 4\pi$ but rather uncertain~\cite{amsler,dm2_rhorho}.
We thus do not calculate this process in this paper.

\subsection{Radiative $J/\psi$ decay rate formula}

The partial decay width for 
a radiative $J/\psi$ decay, $J/\psi\to \gamma(abc)$,
is given by
\begin{eqnarray}
d\Gamma_{J/\psi\to \gamma(abc)}
&=& {{\cal B}\over 2^5 (2\pi)^8 m_{J/\psi}} |{\cal M}_{J/\psi\to \gamma(abc)}|^2
\delta^{(4)}(p_i-p_f) \nonumber\\
&&\times {d^3p_a\over E_a}
 {d^3p_b\over E_b}
 {d^3p_c\over E_c}
 {d^3p_\gamma\over E_\gamma}
\nonumber\\
 &=& 
{{\cal B}\over (2\pi)^5} 
 {\tilde{p}^2_\gamma \over \tilde{E}_\gamma}
{ |{\cal M}_{J/\psi\to \gamma(abc)}|^2 
\over 32\, m_{J/\psi}\, E^2} 
d \tilde{p}_\gamma\, d m^2_{ab}\, d m^2_{ac} ,
\nonumber\\
\label{eq:decay-formula1}
\end{eqnarray}
where $m_{ab}$ and $m_{ac}$ are the invariant masses of the
 $ab$ and $ac$ subsystems, respectively; 
$\tilde{p}_\gamma$ denotes the photon momentum in the $J/\psi$-at-rest frame;
${\cal M}_{J/\psi\to \gamma(abc)}$ is the invariant amplitude that is
 related to Eq.~(\ref{eq:amp_a_sum}) with an overall kinematical factor. 
A Bose factor ${\cal B}$ is: 
${\cal B}=1/3!$ for identical three particles $abc$;
${\cal B}=1/2!$ for identical two particles among $abc$;
${\cal B}=1$ otherwise.
The $J/\psi$ spin state is implicitly averaged.

Using our amplitude of Eq.~(\ref{eq:amp_a})
for the $J/\psi$ radiative decays via $M^*$-excitations,
the decay formula of Eq.~(\ref{eq:decay-formula1}) can be written as
\begin{eqnarray}
{d\Gamma_{J/\psi\to \gamma(abc)}\over  d E}
 &=& 
\sum_{J^{PC}}
{d\Gamma^{J^{PC}}_{J/\psi\to \gamma(abc)}\over  d E}
\label{eq:decay-formula2-sum}
\end{eqnarray}
with
\begin{eqnarray}
{d\Gamma^{J^{PC}}_{J/\psi\to\gamma(abc)}\over  d E}
 &=& 
{2 E^2\over \pi} 
\sum_{ij s^z_{M^*}}
\sum_{k\ell s^{\prime z}_{M^*}}
d\Gamma^{ik}_{M^*\to abc}
{[\bar{G}_{M^*}(E)]_{ij}
\over \sqrt{m_{M^*_i}m_{M^*_j}}} 
\nonumber\\
&&\times 
 {[\bar{G}_{M^*}(E)]^*_{k\ell}
\over \sqrt{m_{M^*_k}m_{M^*_\ell}
}} 
\Gamma^{j\ell}_{J/\psi\to M^*\gamma}  \ ,
\label{eq:decay-formula2}
\end{eqnarray}
where 
$s^z_{M^*}$ ($s^{\prime z}_{M^*}$) is a spin state of $M^*$
in $\bar{G}_{M^*}$ ($[\bar{G}_{M^*}]^*$), 
and
\begin{eqnarray}
\label{eq:decay-formula3}
d\Gamma^{ij}_{M^*\to abc}
&\equiv& {\cal B}
{{\cal M}_{M^*_i\to abc}\,
{\cal M}_{M^*_j\to abc}^*
\over (2\pi)^3\, 32\, E^3}
d m^2_{ab}\, d m^2_{ac} \ , \\
\Gamma^{ij}_{J/\psi\to M^*\gamma}
&\equiv& 
{1\over 8\pi}
{\tilde{p}_\gamma\over m^2_{J/\psi}}
{\cal M}_{J/\psi\to M^*_i\gamma}
{\cal M}_{J/\psi\to M^*_j\gamma}^* \ .
\label{eq:decay-formula4}
\end{eqnarray}
The invariant amplitudes 
${\cal M}_{M^*_i\to abc}$ and 
${\cal M}_{J/\psi\to M^*_j\gamma}$
are related to components of 
the amplitude in Eq.~(\ref{eq:amp_a}) by
\begin{eqnarray}
{\cal M}_{M^*_i\to abc}
&=& 
 - (2\pi)^3 \sqrt{16 m_{M^*_i} E_a E_b E_c}\, T_{M^*_i\to abc} ,
\label{eq:invamp}
\end{eqnarray}
with 
\begin{eqnarray}
T_{M^*_i\to abc} &=& 
\sum^{\rm cyclic}_{abc}
\sum_{RR's_R^z}
\Gamma_{ab,R}\,
\tau_{R,R'}(p_c,E-E_{c})\,
\nonumber\\
&&\times \bar{\Gamma}_{cR',M^*_i}(\bm{p}_c, E) ,
\end{eqnarray}
and 
\begin{eqnarray}
{\cal M}_{J/\psi\to M^*_j\gamma}
&=& 
\sqrt{8 \tilde{E}_\gamma\, m_{J/\psi}\, m_{M^*_j}}\, 
\Gamma_{\gamma M^*_j,J/\psi} \ .
\end{eqnarray}
For the case of $i=j$,
Eqs.~(\ref{eq:decay-formula3}) and (\ref{eq:decay-formula4})
reduce to the standard formulas of the $M^*$-decay Dalitz plot distribution 
and $J/\psi$ two-body decay width, respectively.
Our decay formula of Eq.~(\ref{eq:decay-formula2}) 
can be made look similar to that of a Breit-Wigner model by 
a replacement:
$[\bar{G}_{M^*}(E)]_{ij} \to \delta_{ij}/(E - M_{M^*_i} + i \Gamma_{M^*_i}/2)$,
with $M_{M^*_i}$ and $\Gamma_{M^*_i}$
being Breit-Wigner mass and width, respectively.

\section{Data analysis and
 $\eta(1405/1475)$ poles
}
\label{sec:analysis}

In this paper, we study
radiative $J/\psi$ decays via $\eta(1405/1475)$ excitations
with the unitary coupled-channel model described above.
We thus consider only the $J^{PC}=0^{-+}$ partial wave contribution 
in the above formulas. 
In the following, we discuss our dataset, 
our default setup of the model, and analysis results.

\subsection{Dataset}
A main part of our dataset is $K_SK_S\pi^0$ Dalitz plot pseudodata.
We generate the pseudodata using
the $E$-dependent $0^{-+}$ partial wave amplitude
from the recent BESIII Monte Carlo (MC) analysis on $J/\psi\to\gamma(K_SK_S\pi^0)$~\cite{bes3_mc}.
We often denote this process by 
$J/\psi\to\gamma (0^{-+})\to\gamma(K_SK_S\pi^0)$.
The pseudodata is therefore
detection efficiency-corrected and background-free.

The pseudodata includes $\sim 1.23 \times 10^5$ events in total, being consistent
with the BESIII data, and is binned as follows.
The range of $1300\le E\le 1600$~MeV is divided into 
30 $E$ bins (10~MeV bin width; labeled by $l$).
Furthermore, in each $E$ bin,
we equally divide
$(0.95\, {\rm GeV})^2\le m^2_{K_SK_S}\le (1.50\, {\rm GeV})^2$ and 
$(0.60\, {\rm GeV})^2\le m^2_{K_S\pi^0}\le (1.15\, {\rm GeV})^2$ 
into $50\times 50$ bins (labeled by $m$); $m_{ab}$ is the $ab$ invariant mass.
We denote
the event numbers in
$\{l,m\}$ and $l$-th 
bins by
$N_{l,m}$ and $\bar N_{l}(\equiv\sum_m N_{l,m})$, respectively;
their statistical uncertainties are
$\sqrt{N_{l,m}}$ and $\sqrt{\bar N_{l}}$, respectively.
We fit both $\{N_{l,m}\}$ and $\{\bar N_{l}\}$ pseudodata,
since $\{N_{l,m}\}$ and $\{\bar N_{l}\}$ would efficiently constrain
the detailed decay dynamics and
the resonant behavior (pole structure) of $\eta(1405/1475)$, respectively.
We use the bootstrap method~\cite{bootstrap} to
estimate the statistical uncertainty of the model, 
and we thus generate and fit 50 pseudodata samples.

Other final states from the radiative $J/\psi$ decays
are also considered in our analysis.
We fit the model to a ratio of partial decay widths~\cite{pdg}
\begin{eqnarray}
R_1^{\rm exp}
&&={
\Gamma  [J/\psi\to\gamma\eta(1405/1475)\to\gamma(K\bar{K}\pi)]
\over 
\Gamma  [J/\psi\to\gamma\eta(1405/1475)\to\gamma(\eta\pi^+\pi^-)]
}
\nonumber\\
&&=
{ (2.8\pm 0.6)\times 10^{-3} \over
(3.0\pm 0.5)\times 10^{-4}
}
= 6.8-11.9 \ ,
\label{eq:R1exp}
\end{eqnarray}
and also another ratio~\cite{mark3_rhog,bes2_rhog}
\begin{eqnarray}
R_2^{\rm exp}
&&={
\Gamma  [J/\psi\to\gamma\eta(1405/1475)\to\gamma(\rho^0\gamma)]
\over 
\Gamma  [J/\psi\to\gamma\eta(1405/1475)\to\gamma(K\bar{K}\pi)]
}
\nonumber\\
&&= 0.015-0.043 \ .
\label{eq:R2exp}
\end{eqnarray}
The partial widths $\Gamma$ in the above ratios are calculated
by integrating the $E$
distributions [Eq.~(\ref{eq:decay-formula2-sum})]
for $K\bar{K}\pi$, $\pi^+\pi^-\eta$, and $\rho^0\gamma$
final states over the range of $1350$~MeV $<E< 1550$~MeV.
The ratio of Eq.~(\ref{eq:R1exp}) is important to constrain
the $a_0(980)\pi$ contributions 
since the relative coupling strengths of
$a_0(980)\to K\bar{K}$ and $a_0(980)\to \eta\pi$ 
are experimentally fixed in a certain range~\cite{a0_980_ppbar,bes3_a0,omeg_a0,a0_980_gg}.
Also, $f_0\eta$ and $\rho\rho$ channels indirectly
contribute to $K_SK_S\pi^0$ through loops,
and therefore the $K_SK_S\pi^0$ data does not constrain
their parameters well.
Since these channels
directly contribute to the $\eta\pi^+\pi^-$ and $\rho^0\gamma$
final states, 
the above ratios will be a good constraint.
The partial width for all $K\bar{K}\pi$ final states 
in Eqs.~(\ref{eq:R1exp}) and (\ref{eq:R2exp}) 
is 12 times larger than that of $K_SK_S\pi^0$, as determined by the isospin CG coefficients. 

The MC solution-based
$\{N_{l,m}\}$, $\{\bar N_{l}\}$, $R_1^{\rm exp}$, and $R_2^{\rm exp}$
are simultaneously fitted, with a $\chi^2$-minimization,
by our default model described in the next subsection;
the actual BESIII data are not directly fitted.
We calculate $\chi^2$ from $\{N_{l,m}\}$
by comparing $N_{l,m}$ to
the differential decay width 
[$d\Gamma_{J/\psi\to \gamma(abc)}/dE dm^2_{ab} dm^2_{ac}$
of Eqs.~(\ref{eq:decay-formula2-sum})--(\ref{eq:decay-formula4})]
evaluated at the bin center and multiplied by the bin volume. 
We omit $N_{l,m}$ on the phase-space boundary from
the $\chi^2$ calculation. 
This simplified procedure keeps the computation time reasonable.
Also, if a bin has $N_{l,m}<10$, it is combined with neighboring bins
to have more than 9 events for the $\chi^2$ calculation.
The number of bins for $\{N_{l,m}\}$
depends on the pseudodata samples, and is 4496--4575.
$\chi^2$ from $\{\bar N_{l}\}$, $R_1^{\rm exp}$, and $R_2^{\rm exp}$
are weighted appropriately to reasonably constrain the model.

\subsection{Model setup}
\label{subsec:model}

For the present analysis of the radiative $J/\psi$ decays, 
we consider the following coupled-channels as a default
in our model described in Sec.~\ref{sec:model}.
We include two bare $M^*$ of $J^{PC}=0^{-+}$; we refer to them as
$\eta^*$ hereafter.
The $Rc$ channels are
$Rc=K^*(892)\bar K$, $\kappa\bar K$,
$a_0(980)\pi$, $a_2(1320)\pi, f_0\eta, \rho(770)\rho(770)$, 
and $f_0\pi$, where charge indices are suppressed~\footnote{
$\kappa$ is also referred to as $K^*_0(700)$ in the literature.
}.
To form positive $C$-parity states,
$\bar{K}^*(892) K$ and $\bar{\kappa} K$ channels 
are implicitly included.
A symbol $R$ may refer to more than one bare state and/or contact interactions.
For example, 
the $f_0\pi$ channel includes two bare states and one contact interaction
that nonperturbatively couple with
$\pi\pi-K\bar{K}$ continuum states, forming
$f_0(500)$, $f_0(980)$, and $f_0(1370)$ poles; see 
Appendix~A for details.

Regarding 
the $\pi\eta-K\bar{K}$ coupled-channel $s$-wave scattering amplitude that includes an 
$a_0(980)$ pole, we consider two experimental inputs; see 
Appendix~\ref{sec:app-pipi} for details.
First, 
the $a_0(980)$ amplitude 
from the BESIII amplitude analysis on $\chi_{c1}\to\eta\pi^+\pi^-$
constrains the energy dependence of 
our $a_0(980)$ model.
Second, we determine $a_0(980)\to K\bar{K}$ and $a_0(980)\to \eta\pi$
decay strengths using an analysis of
$p\bar{p}\to K\bar{K}\pi, \eta\pi\pi$~\cite{a0_980_ppbar} giving 
$|g_{a_0(980)\to K\bar{K}}/g_{a_0(980)\to \eta\pi}|\sim 1$;
$g_{a_0(980)\to ab}$ is the residue of $a_0(980)\to ab$ decay.
The ratio of branching fractions of
$a_0(980)\to K\bar{K}$ and $a_0(980)\to \eta\pi$,
which can be translated into 
$|g_{a_0(980)\to K\bar{K}}/g_{a_0(980)\to \eta\pi}|$,
has not been precisely determined experimentally~\cite{pdg,a0_980_ppbar,bes3_a0,omeg_a0,a0_980_gg}.
We will discuss later 
possible impacts of using a different $a_0(980)$ model with different
$|g_{a_0(980)\to K\bar{K}}/g_{a_0(980)\to \eta\pi}|$.

We mention the channels considered 
in the BESIII amplitude analysis of $J/\psi\to\gamma(K_SK_S\pi^0)$~\cite{bes3_mc}.
In the $0^{-+}$ partial wave, 
the BESIII considered 
$\eta(1405)$ and $\eta(1475)$ resonances that decay into 
$K^*(892)\bar K$, 
$a_0(980)\pi$, and $a_2(1320)\pi$.
All resonances, except for $a_0(980)$, are
described with Breit-Wigner amplitudes.
No rescattering nor channel-coupling
such as those in the second term of the r.h.s. of 
Fig.~\ref{fig:diag}(c) is taken into account. 
In addition, a nonresonant $K\pi$ $p$-wave 
amplitude supplements the $K^*(892)$ tail region. 
Clearly, our coupled-channel model includes more channels than the
BESIII model does. 
This is to satisfy the coupled-channel three-body unitarity, and also to describe
different final states in a unified manner.

We consider isospin-conserving $\eta^*\to Rc$ decays in Eq.~(\ref{eq:bare_mstar})
for all bare $\eta^*$ and $Rc$ states
specified in the first paragraph of this subsection.
One exception applies to the lighter bare $\eta^*\to\rho\rho$ which is
set to zero.
This is because the lighter bare $\eta^*$ seems consistent with an
excited $s\bar{s}$ state from the quark model~\cite{barnes1997} and
LQCD prediction~\cite{dudek2013}, and
$s\bar{s}\to\rho\rho$ should be small for the OZI rule.

We may add nonresonant (NR) amplitudes 
$A^{J^{PC},{\rm NR}}_{\gamma abc,J/\psi}$,
which does not involve $M^*$ excitations, 
to the resonant amplitudes $A^{J^{PC}}_{\gamma abc,J/\psi}$ 
of Eq.~(\ref{eq:amp_a_sum}).
We can derive
$A^{J^{PC},{\rm NR}}_{\gamma abc,J/\psi}$
and modify 
$A^{J^{PC}}_{\gamma abc,J/\psi}$ 
so that their sum still maintains the three-body unitarity.
However, this introduces too many fitting parameters to determine with 
the dataset in the present analysis.
We thus use a simplified NR amplitude in this work [cf. Eq.~(\ref{eq:invamp2})]:
\begin{eqnarray}
{\cal M}^{\rm NR}_{J/\psi\to \gamma(abc)} = c_{\rm NR}\;
(\tilde{\bm{\epsilon}}_{J/\psi}\times \tilde{\bm{\epsilon}}_{\gamma})\cdot \tilde{\bm{p}}_\gamma,
\label{eq:cnr}
\end{eqnarray}
where $c_{\rm NR}$ is a complex constant.
Only when fitting 
the $J/\psi\to\gamma (0^{-+})\to\gamma K_SK_S\pi^0$
Dalitz plot pseudodata,
this NR term is added 
to ${\cal M}_{J/\psi\to \gamma(abc)}$ in Eq.~(\ref{eq:decay-formula1})
and $c_{\rm NR}$ is determined by the fit. 

\begin{figure*}
\begin{center}
\includegraphics[width=1\textwidth]{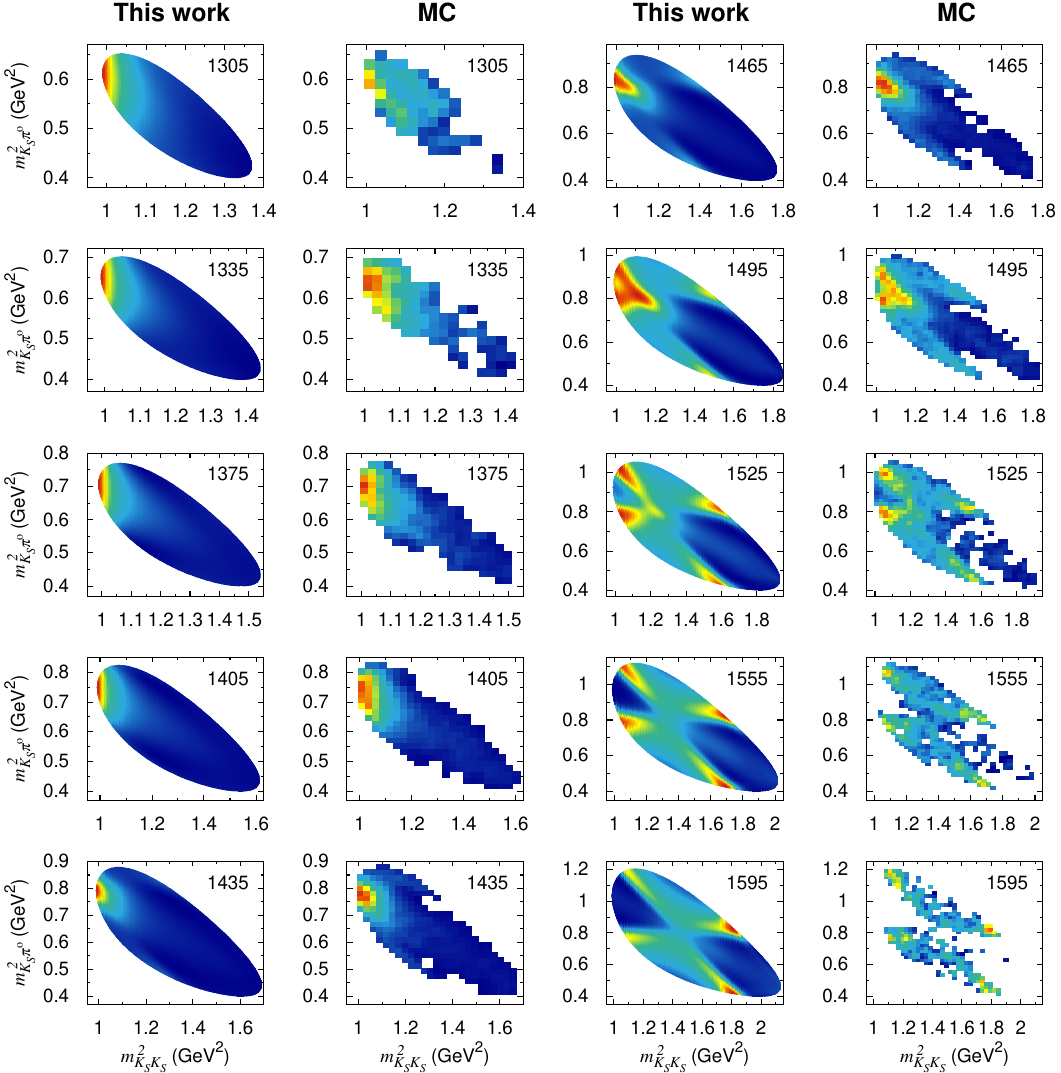}
\end{center}
 \caption{
The $K_SK_S\pi^0$ Dalitz plot distributions for
$J/\psi\to\gamma (0^{-+})\to \gamma(K_SK_S\pi^0)$.
Our fit result and pseudodata (MC) are shown.
The $E$ values used in our calculation 
(the central values of the $E$ bins of the pseudodata)
are indicated in each panel.
The distributions are shown, in the descending order, by the
red, yellow, green, and blue colors.
Depending on $E$, the same color means different absolute values.
 }
\label{fig:fit1}
\end{figure*}

\begin{figure*}
\begin{center}
\includegraphics[width=.8\textwidth]{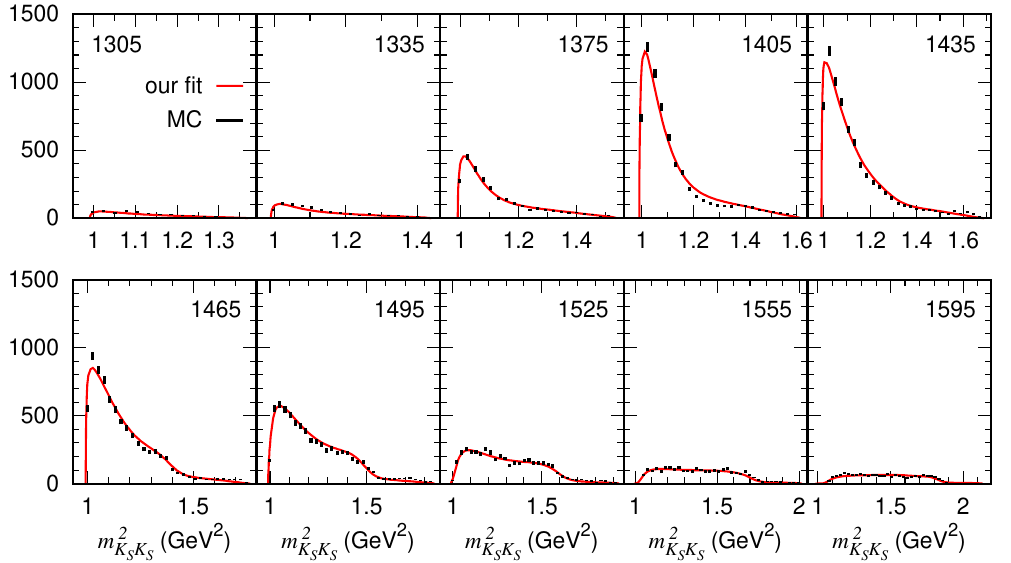}
\end{center}
 \caption{
The $K_SK_S$ invariant mass distributions
for $J/\psi\to\gamma (0^{-+})\to\gamma(K_SK_S\pi^0)$.
Our fit result (red solid curve) and the pseudodata (black error bars) are shown.
The $E$ values are indicated in each panel.
 }
\label{fig:KK-Kpi}
\end{figure*}
\begin{figure*}
\begin{center}
\includegraphics[width=.8\textwidth]{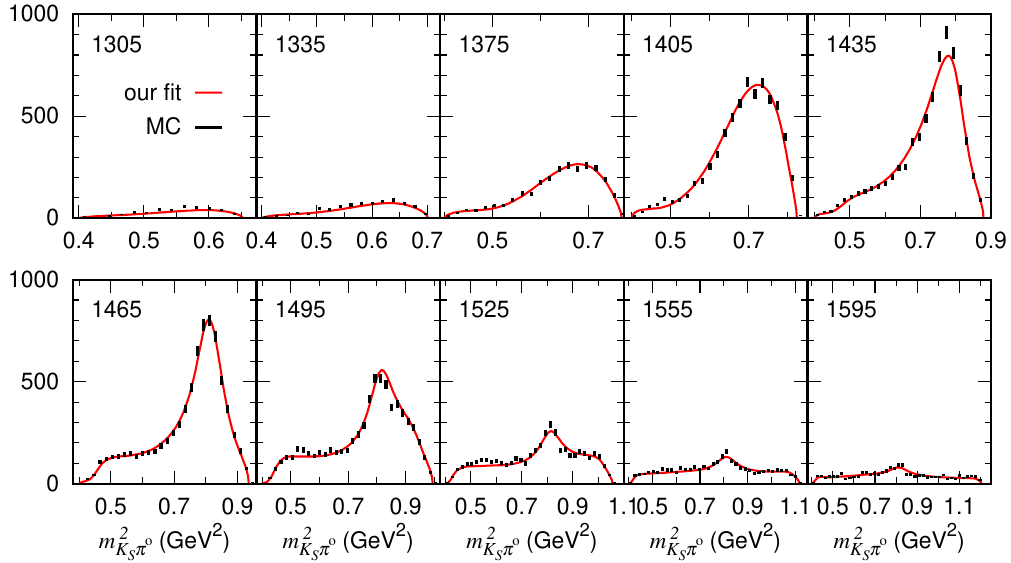}
\end{center}
 \caption{
The $K_S\pi^0$ invariant mass distributions.
Other features are the same as those in 
Fig.~\ref{fig:KK-Kpi}.
 }
\label{fig:KK-Kpi2}
\end{figure*}

\begin{figure*}
\begin{center}
\includegraphics[width=1\textwidth]{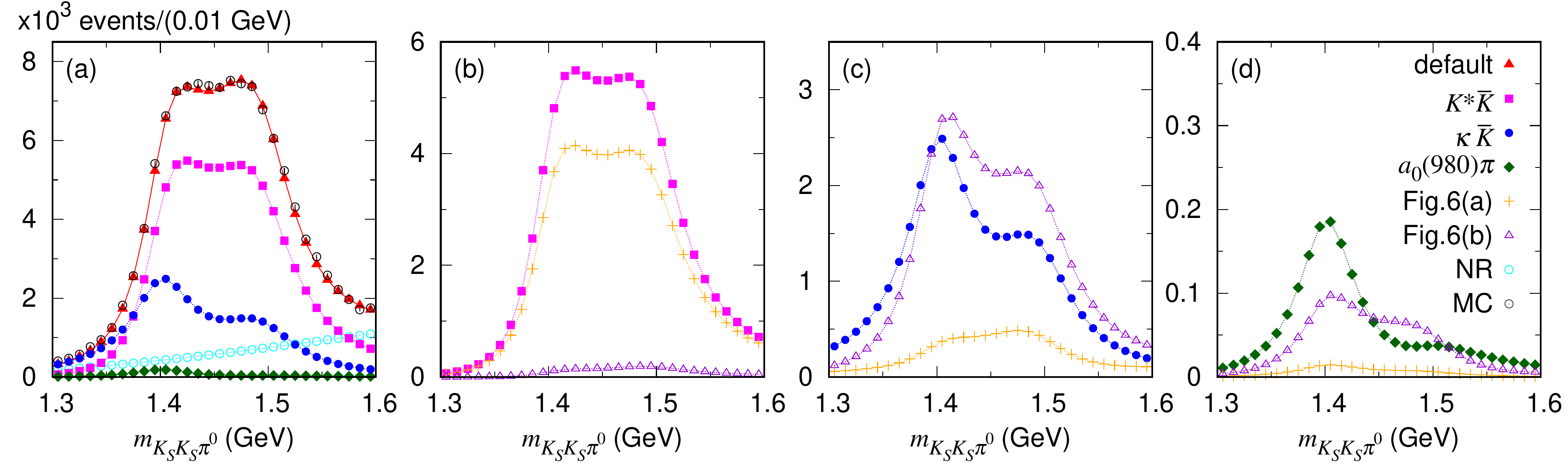}
\end{center}
 \caption{The $m_{K_SK_S\pi^0}(=E)$ distribution for 
$J/\psi\to\gamma (0^{-+})\to\gamma(K_SK_S\pi^0)$.
(a) Our default fit to the pseudodata (MC).
Final $K^*\bar{K}$, $\kappa\bar{K}$, and $a_0(980)\pi$ contributions as
 well as the nonresonant (NR) contribution
 are also shown.
(b) The final $K^*\bar{K}$ contribution.
The main contributions from the diagrams of Figs.~\ref{fig:diag2}(a) and
 \ref{fig:diag2}(b) are also shown.
(c) [(d)] The final $\kappa\bar{K}$ [$a_0(980)\pi$] contribution
shown similarly to the panel (b).
Lines connecting the points are just for guiding eyes.
Figure~\ref{fig:kkpi}(a) is taken from Ref.~\cite{letter}.
Copyright (2023) APS.
 }
\label{fig:kkpi}
\end{figure*}
\begin{figure}
\begin{center}
\includegraphics[width=.48\textwidth]{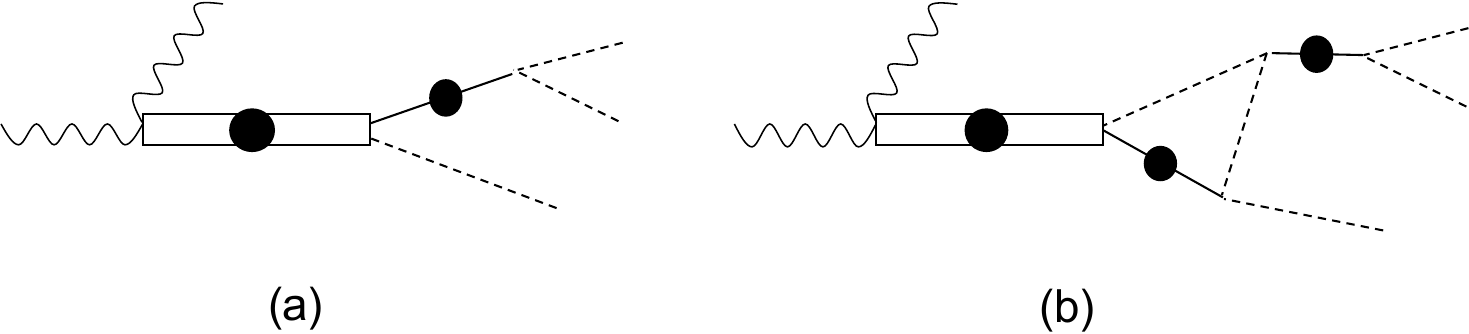}
\end{center}
 \caption{Main $\eta^*$ decay mechanisms
included in Fig.~\ref{fig:diag}(a):
(a) direct decays; (b) single rescattering due to $V$ of Eq.~(\ref{eq:vvv}).
 }
\label{fig:diag2}
\end{figure}

\begin{figure}
\begin{center}
\includegraphics[width=.48\textwidth]{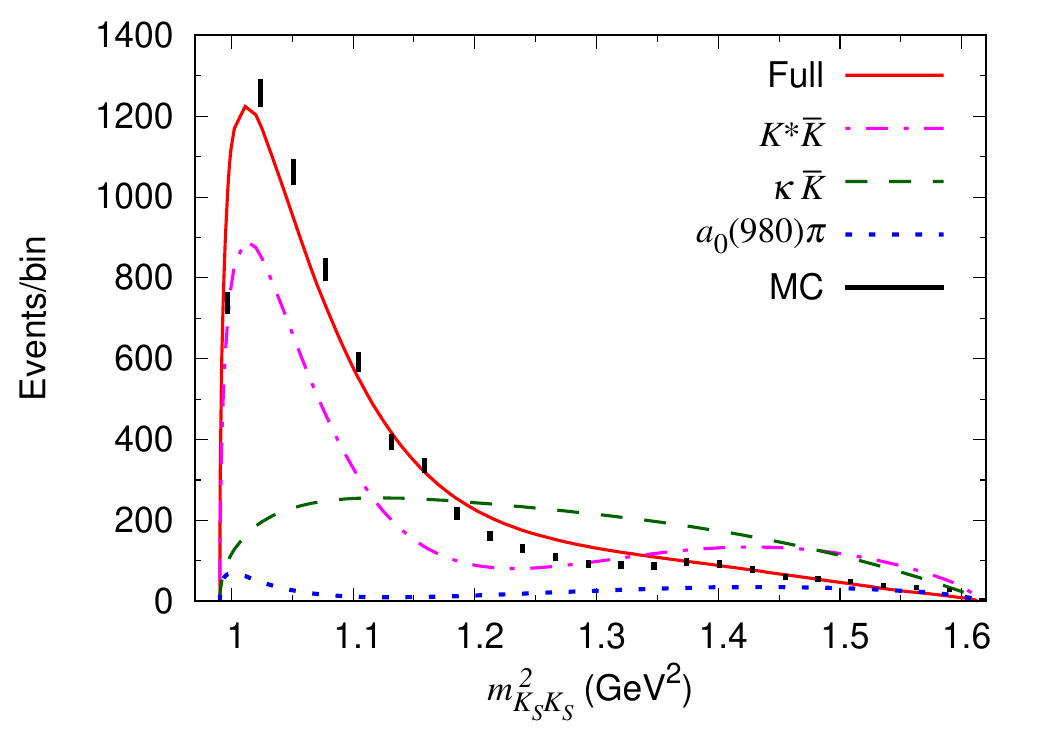}
\end{center}
 \caption{
Different final $Rc$ contributions to
the $K_SK_S$ invariant mass distribution at $E=1405$~MeV.
 }
\label{fig:a0-like}
\end{figure}

We summarize the parameters 
fitted to the dataset discussed in the previous subsection.
We have two bare $\eta^*$ masses in Eq.~(\ref{eq:mstar-g1}),
two complex coupling constants ($g_{J/\psi \eta^*_j\gamma}$)
in Eq.~(\ref{eq:invamp2}),
and one complex constant $c_{\rm NR}$ in Eq.~(\ref{eq:cnr}).
We also adjust real coupling parameters $C^{M^*_i}_{(c R)_l}$
in Eq.~(\ref{eq:bare_mstar}).
While the cutoffs $\Lambda^{M^*_i}_{(c R)_l}$ in Eq.~(\ref{eq:bare_mstar})
are also adjustable,
we fix them to 700~MeV in this work to reduce the number of fitting
parameters and speed up the fitting procedure. 
Since the overall strength and phase of the full amplitude are
arbitrary, we have 25 fitting parameters in total.
The parameter values obtained from the fit are presented in 
Table~\ref{tab:param} of Appendix~\ref{app2}.

All the radiative $J/\psi$ decay processes included 
in our dataset for the fit are isospin-conserving.
Since the isospin-violating effects are very small in these processes, 
we make the model isospin-symmetric for fitting and extracting poles and thus
use the averaged mass for each isospin multiplet. 
The amplitude formulas in Sec.~\ref{sec:model-1}
reduce to isospin symmetric ones given in Refs.~\cite{3pi,d-decay}.
This simplification significantly speeds up 
the fitting and pole extraction procedures.
When calculating the isospin-violating $J/\psi\to\gamma (0^{-+})\to\gamma(\pi\pi\pi)$
amplitude of Eq.~(\ref{eq:amp_a}),
we still use the isospin symmetric
$\bar{G}_{\eta^*}(E)$ and parameters determined by fitting the
dataset, and the pole positions stay the same;
the isospin violations occur in 
$\tau_{R,R'}$ and $\bar{\Gamma}_{cR,\eta^*_i}$
due to the difference between $m_{K^\pm}$ and $m_{K^0}$.

\subsection{Fits to $K_SK_S\pi^0$ Dalitz plot pseudodata generated from
  BESIII $0^{-+}$ amplitude}

By fitting the 50 bootstrap samples of the $K_SK_S\pi^0$ Dalitz plot pseudodata
with the default dynamical contents described above,
we obtain
$\chi^2/{\rm ndf}=$ 1.40--1.54 (ndf: number of degrees of freedom)
by comparing with $\{N_{l,m}\}$.
The ratios of 
Eqs.~(\ref{eq:R1exp}) and (\ref{eq:R2exp})
are also fitted simultaneously, obtaining 
$R_1^{\rm th}\sim 7.5$ and $R_2^{\rm th}\sim 0.025$, respectively.

The Dalitz plot distributions obtained from 
the fit are shown in Fig.~\ref{fig:fit1}
for representative $E$ values,
in comparison with one of the bootstrap samples~\footnote{
The same bootstrap sample is also shown in 
Figs~\ref{fig:KK-Kpi}, \ref{fig:KK-Kpi2}, \ref{fig:kkpi}(a), \ref{fig:a0-like},
and \ref{fig:kkpi-one-bare}.}.
The fit quality is reasonable overall.
For $E\ltap 1400$~MeV, there is a peak near the $K_SK_S$ threshold.
While this is seemingly the $a_0(980)$ contribution, 
it is actually due to a constructive interference between $K^*(892)$ and $\bar K^*(892)$,
as detailed later.
For $E\gtap 1430$~MeV, on the other hand, 
the main pattern is mostly understood as the $K^*(892)$ and $\bar K^*(892)$
resonance contributions.
The good fit quality can be seen more clearly in the  
$K_SK_S$ and $K_S\pi^0$ invariant mass distributions 
as shown in Figs.~\ref{fig:KK-Kpi} and \ref{fig:KK-Kpi2}, 
respectively.
The model is well-fitted to 
the $K^*$ peak
(the sharp peak near the $K_SK_S$ threshold) in the 
$m^2_{K_S\pi^0}$ 
($m^2_{K_SK_S}$) distributions.

The $E$ dependence of
the radiative $J/\psi$ decay to $K_SK_S\pi^0$,
obtained by integrating the Dalitz plots in Fig.~\ref{fig:fit1},
is shown in Fig.~\ref{fig:kkpi}(a).
The $E$-dependence would be largely determined by the pole structure
of the $\eta(1405/1475)$ resonances. 
The $E$ distribution shows a broad peak with an almost flat top, and
our model reasonably agrees with the pseudodata. 
We now study dynamical details.
The $\eta^*$ decay mechanisms can be separated according to 
$Rc$ states in Fig.~\ref{fig:diag}(a) 
that directly couple to the final states.
We will refer to these $Rc$ states as {\it final} $Rc$ states.
Contributions from the final
$K^*\bar{K}$, 
$\kappa\bar{K}$, 
and $a_0(980)\pi$ states 
are shown separately in 
Fig.~\ref{fig:kkpi}(a).
The final $K^*\bar{K}$ and $\kappa\bar{K}$ 
contributions are the first and second largest, 
while the final $a_0(980)\pi$ contribution is very small.
The constant nonresonant contribution from Eq.~(\ref{eq:cnr})
gives a small phase-space shape contribution. 

The final $K^*\bar{K}$, $\kappa\bar{K}$, and $a_0(980)\pi$ contributions
are also shown separately in 
Figs.~\ref{fig:kkpi}(b), \ref{fig:kkpi}(c), and \ref{fig:kkpi}(d),
respectively, and main contributions from the diagrams in
Fig.~\ref{fig:diag2} are also shown. 
The direct decays of Fig.~\ref{fig:diag2}(a) and 
single-rescattering mechanisms of Fig.~\ref{fig:diag2}(b) 
are obtained by perturbatively expanding the dressed 
$\eta^*$ decay vertex of Fig.~\ref{fig:diag}(c) in terms of 
 $V$ in Eq.~(\ref{eq:vvv}),
and taking the first two terms.
The final $K^*\bar{K}$ contribution is mostly from the direct decay, while
the final $\kappa\bar{K}$ and $a_0(980)\pi$ contributions are
dominantly from the single-rescattering mechanism  
and therefore a coupled-channel effect.
The $K^*\bar{K} K$ triangle loop causes a triangle singularity (TS) in the 
the final $a_0(980)\pi$ contribution at $E\sim 1.4$~GeV.
However, we do not find a large contribution from the TS.
The TS-induced enhancement may have been suppressed since 
the $K^*\bar K$ pair is relatively $p$-wave.

Figure~\ref{fig:a0-like} illustrates the mechanism that creates the sharp 
$a_0(980)$-like
enhancement near the $K_SK_S$ threshold.
Clearly, the final $K^*\bar{K}$ contribution alone creates the structure
mostly, and the other mechanisms moderately change it.
The final $a_0(980)\pi$ contribution is minor. 
As the Dalitz plots in Fig.~\ref{fig:fit1} show, 
$K^*$ and $\bar K^*$ constructively interfere to generate a peak near the 
$K_SK_S$ threshold for $E=1.45$--1.5~GeV.
The $a_0(980)$-like peaks seen in $E=1.3$--1.45~GeV are also caused by the
same mechanism.

The BESIII model obtained from their amplitude analysis
describes the data rather differently from ours 
(see Fig.~3 of Ref.~\cite{bes3_mc}) such as:
(i) The $a_0(980)\pi$ contribution is the largest overall;
(ii) The $K^*\bar{K}$ contribution
is comparable to $a_0(980)\pi$ only around $E=1.5$~GeV;
(iii) The $\kappa\bar{K}$ channel is not included.
These differences come mainly from the fact that
our model is fitted not only to the $K_SK_S\pi^0$ Dalitz plot pseudodata
but also to the ratios of Eqs.~(\ref{eq:R1exp}) and (\ref{eq:R2exp});
the BESIII model was fitted to the $J/\psi\to\gamma(K_SK_S\pi^0)$ data only.
The ratio of Eq.~(\ref{eq:R1exp}) is, albeit a large uncertainty, an important
constraint on the final $a_0(980)\pi$ contribution to $\eta^*\to K\bar{K}\pi$,
since the relative coupling of $a_0(980)\to K\bar{K}$ to 
$a_0(980)\to \pi\eta$ is experimentally determined in a certain
range~\cite{pdg,a0_980_ppbar,bes3_a0,omeg_a0,a0_980_gg}.
The final $a_0(980)\pi$ contribution to $K\bar{K}\pi$ needs to be small
as in our model in order to satisfy the ratio of 
Eq.~(\ref{eq:R1exp}).
Furthermore, the $\kappa\bar{K}$ channel in our model gives a
substantial contribution through the channel-coupling required by the unitarity.

Since the $a_0(980)\pi$ contribution is very different between our and
the BESIII models, 
one may wonder how much our result depends on a particular $a_0(980)$
model.
As we discussed in Sec.~\ref{subsec:model}, 
our default $a_0(980)$ model is based on Ref.~\cite{a0_980_ppbar} 
and $|g_{a_0(980)\to K\bar{K}}/g_{a_0(980)\to \eta\pi}|\sim 1$.
In PDG~\cite{pdg}, two other analyses of Refs.~\cite{bes3_a0,omeg_a0}
were considered in averaging 
${\cal B}(a_0(980)\to \eta\pi)/{\cal B}(a_0(980)\to K\bar{K})$.
This ratio of the branchings can be translated into
$|g_{a_0(980)\to K\bar{K}}/g_{a_0(980)\to \eta\pi}|\sim 0.77$~\cite{bes3_a0}
and 
$|g_{a_0(980)\to K\bar{K}}/g_{a_0(980)\to \eta\pi}|\sim 0.85$~\cite{omeg_a0}.
Thus if we use an $a_0(980)$ model based on Refs.~\cite{bes3_a0,omeg_a0}
in our present analysis, 
the corresponding $a_0(980)\pi$ contribution would be even smaller. 
There is also an analysis on 
$\gamma\gamma\to K\bar{K}, \eta\pi$ giving 
$|g_{a_0(980)\to K\bar{K}}/g_{a_0(980)\to \eta\pi}|\sim 2$~\cite{a0_980_gg}.
However, this analysis 
did not include $a_0(980)\to K\bar{K}$ data.
Even if we use an $a_0(980)$ model based on this, 
our default result would not qualitatively change since
the $a_0(980)\pi$ contribution could be at most $\sim 4$ times larger than
our default result.

\subsection{Fit with one bare $\eta^*$ state
}
\label{sec:one-bare}

\begin{figure}
\begin{center}
\includegraphics[width=.49\textwidth]{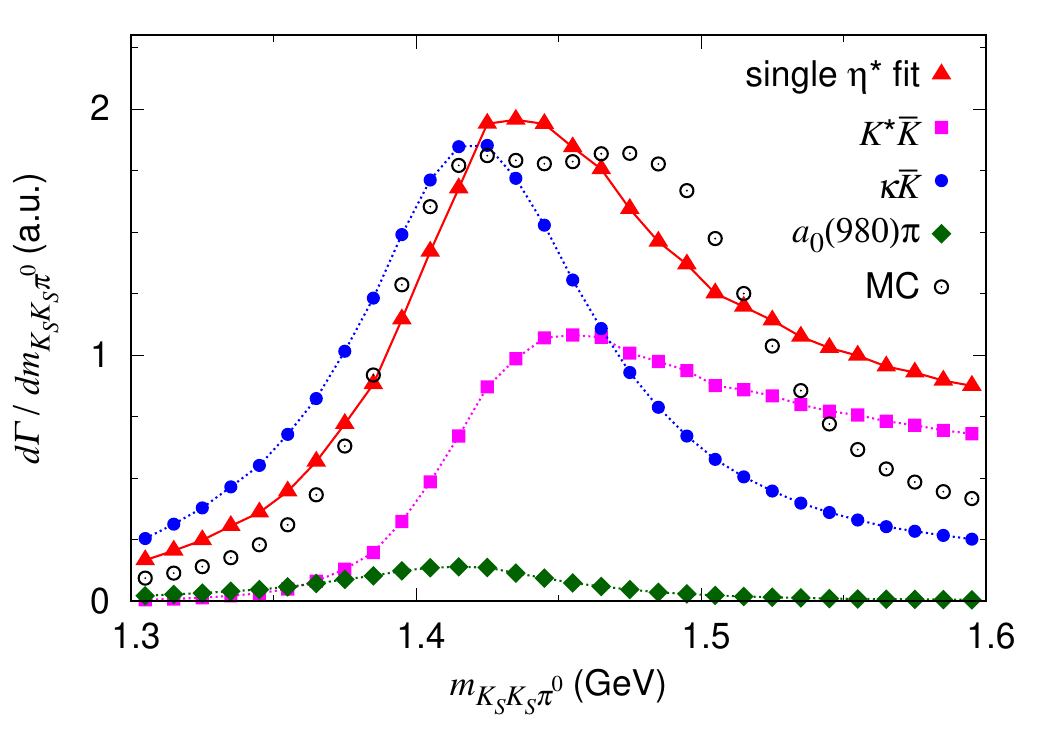}
\end{center}
 \caption{Single bare $\eta^*$ fit (red triangles) to the pseudodata (MC).
Various final $Rc$ contributions are also shown.
 }
\label{fig:kkpi-one-bare}
\end{figure}
It is important to examine if the BESIII data can also be fitted with 
a single bare $\eta^*$ model, since 
the $\eta(1405/1475)$ was claimed to be a single state in the
literature. 
We try to fit only the $m_{K_SK_S\pi^0}(=E)$ distribution,
but a reasonable fit is not achievable.
The result is shown in Fig.~\ref{fig:kkpi-one-bare} along with the final
$Rc$ contributions.
The final $\kappa\bar{K}$ and $a_0\pi$ contributions have lineshapes
expected from the $\eta^*$ pole position, $1416-61i$~MeV.
The triangle singularity caused by the $K^*\bar{K}K$-loop does not
noticeably shift the lineshape of the final $a_0(980)\pi$ contribution.
The lineshape of the final $K^*\bar{K}$ contribution has its peak at
30--40~MeV higher than the peak positions of 
the final $\kappa\bar{K}$ and $a_0(980)\pi$ contributions,
since its threshold opens at $E\sim 1.4$~GeV and 
the $K^*\bar{K}$ pair is relatively $p$-wave.
Still, 
the peak shift is not large enough to explain
the significantly broader peak of the pseudodata.

Another possible single-state solution for $\eta(1405/1475)$ 
describes the BESIII data by including an interference with 
$\eta(1295)$. 
To examine this possibility, we include two bare $\eta^*$ states,
and restrict one of the bare masses below 1.4~GeV, and the other around 
1.6~GeV.
We are not able to obtain a reasonable fit to the pseudodata with this
model. 
We thus conclude that two bare $\eta^*$ for $\eta(1405/1475)$ 
are necessary to reasonably
fit the $K_SK_S\pi^0$ pseudodata generated from the BESIII $0^{-+}$ amplitude.

\subsection{Pole positions for $\eta(1405)$ and $\eta(1475)$}

\begin{table}[b]
\renewcommand{\arraystretch}{1.6}
\tabcolsep=4.mm
\caption{\label{tab:pole}
Locations of poles ($E_{\eta^*}$); each pole is labeled by $\alpha$.
The mass, width and $E_{\eta^*}$ are related by
$M={\rm Re} [E_{\eta^*}]$ and $\Gamma=-2{\rm Im} [E_{\eta^*}]$.
Each pole is located on 
the Riemann sheet (RS) specified by
 $(s_{K^*\bar{K}},s_{a_2(1320)\pi})$;
 $s_{x}=p(u)$ indicates the physical (unphysical) sheet of a channel $x$.
Breit-Wigner parameters from the BESIII analysis are also shown.
Errors are statistical only.
Table taken from Ref.~\cite{letter}. Copyright (2023) APS.
}
\begin{tabular}{lccc}
 & $M$ (MeV) & $\Gamma$ (MeV) & RS \\\hline
$\alpha=1$ &$1401.6\pm 0.6$ & $65.8\pm 1.0$ & $(up)$\\
$\alpha=2$ &$1401.6\pm 0.6$ & $66.3\pm 0.9$ & $(pp)$\\
$\alpha=3$ &$1495.0\pm 1.5$ & $86.4\pm 1.8$ & $(up)$\\\hline
BESIII~\cite{bes3_mc} & 1391.7$\pm 0.7$ & 60.8$\pm 1.2$ &   \\
                      & 1507.6$\pm 1.6$& 115.8$\pm 2.4$ &   \\
\end{tabular}
\end{table}
The properties of a resonance are characterized by its pole position and
residue of the (scattering or decay) amplitude. 
In the present unitary coupled-channel framework, 
a pole position corresponds to a complex energy $E$ that
satisfies ${\rm det}\, [\bar{G}^{-1}(E)]=0$, where 
$\bar{G}^{-1}(E)$ has been defined in Eq.~(\ref{eq:mstar-g1})
and is analytically continued to the complex $E$-plane.
The analytic continuation involves deformations of the integral paths
in Eqs.~(\ref{eq:RR-self}), (\ref{eq:dressed-g}), (\ref{eq:pw-tcr}), and
(\ref{eq:mstar-sigma}).
Otherwise, singularities on the complex momentum planes cross the
real momentum paths as $E$ goes to complex values, invalidating the analytic continuation.
The driving term $Z^{\bar{c},J^{PC}}_{(c'R')_{l'},(cR)_{l}}$ in
Eq.~(\ref{eq:vvv}) 
and 
$\tau_{R,R'}$ in Eqs.~(\ref{eq:dressed-g}), (\ref{eq:pw-tcr}), and
(\ref{eq:mstar-sigma}) cause such singularities.
To avoid these singularities,
a possible deformed path to be used in 
Eqs.~(\ref{eq:dressed-g}), (\ref{eq:pw-tcr}), and (\ref{eq:mstar-sigma})
can be found in Fig.~7 of Ref.~\cite{a1-gwu}.
The energy denominator in Eq.~(\ref{eq:RR-self}) also causes a
singularity and, for a complex $E$, we need to avoid it by choosing a
deformed path as found in Fig.~3 of Ref.~\cite{a1-gwu}.
Our procedure of the analytic continuation is very similar to those
discussed in detail in  Ref.~\cite{a1-gwu},
and we do not go into it further.

\begin{figure}[t]
\begin{center}
\includegraphics[width=.49\textwidth]{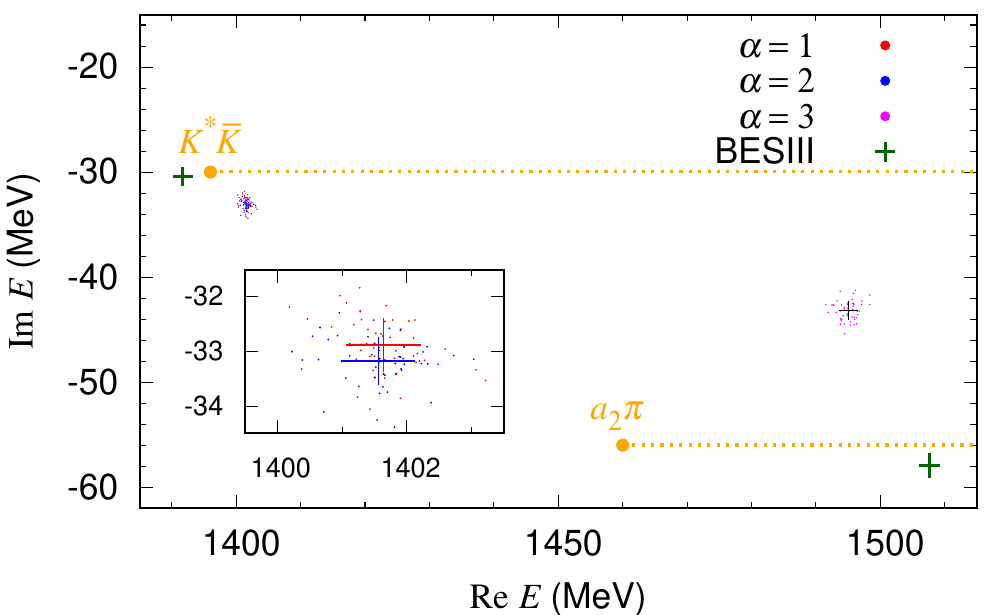}
\end{center}
 \caption{
Locations of $\eta(1405)$ and $\eta(1475)$
poles (labeled by $\alpha$) from 50 bootstrap fits.
Averaged locations of poles and their standard deviations are indicated by
the crosses.
The $K^*(892)\bar{K}$ and $a_2(1320)\pi$ branch points and cuts
are shown by the orange circles and dotted lines, respectively.
The BESIII result~\cite{bes3_mc} (Breit-Wigner parameters) is shown by
 the green crosses. 
The inset shows the $\alpha=1,2$ region.
Figure taken from Ref.~\cite{letter}. Copyright (2023) APS.
 }
\label{fig:pole}
\end{figure}

We search for poles in the region of 
${\rm Re}[E]=1300-1600$~MeV and 
${\rm Im}[-E]=0-200$~MeV
on the relevant Riemann sheets (RS) close to the physical energy. 
We find three poles as listed in Table~\ref{tab:pole}.
The poles are labeled by $\alpha=1,2$ [$\alpha=3$]
corresponding to $\eta(1405)$ [$\eta(1475)$].
The $\eta(1405/1475)$ poles are close to the branch points associated
with the $K^*(892)\bar{K}$ and $a_2(1320)\pi$ thresholds at 
$\sim 1396 - 30i$~MeV and $\sim 1460-56i$~MeV, respectively.
Thus we specify the pole's RS of these channels in Table~\ref{tab:pole};
the relevant RS of the other channels should be clear~\footnote{For the definition of (un)physical sheet,
see the review section 50 ``Resonances'' in Ref.~\cite{pdg}.}.
The locations of the poles and branch points are also shown in Fig.~\ref{fig:pole}.

The BESIII analysis result (Breit-Wigner parameters) is also shown for comparison. 
A noticeable difference is that
our model describes $\eta(1405)$ with two poles ($\alpha=1,2$).
The two pole structure does not mean two physical states but
is simply due to the fact that
a pole coupled to a channel is split into two poles on different RS of 
this channel.
The mass and width values are fairly similar 
between our and the BESIII results.
However, 
the use of the Breit-Wigner amplitude could cause an 
artifact due to the issues discussed in the introduction and below,
which might explain the difference between the two analysis results.
In Ref.~\cite{3pi-2}, a unitary coupled-channel model and an isobar
(Breit-Wigner) model were fitted to the same pseudo-data.
Resonance poles from the two models 
can be significantly different,
particularly when two resonances are overlapping. 
Also, if the pole is located near a threshold,
the lineshape ($E$ dependence)
caused by the pole can be distorted by the branch cut.
In the present case, 
$\eta(1405)$ and $\eta(1475)$ are fairly overlapping
and $\eta(1405)$ is located near the $K^*\bar{K}$ threshold.
Our three-body unitary coupled-channel analysis
fully considers these issues and
is a more appropriate pole-extraction method.

\begin{figure}[t]
\begin{center}
\includegraphics[width=.49\textwidth]{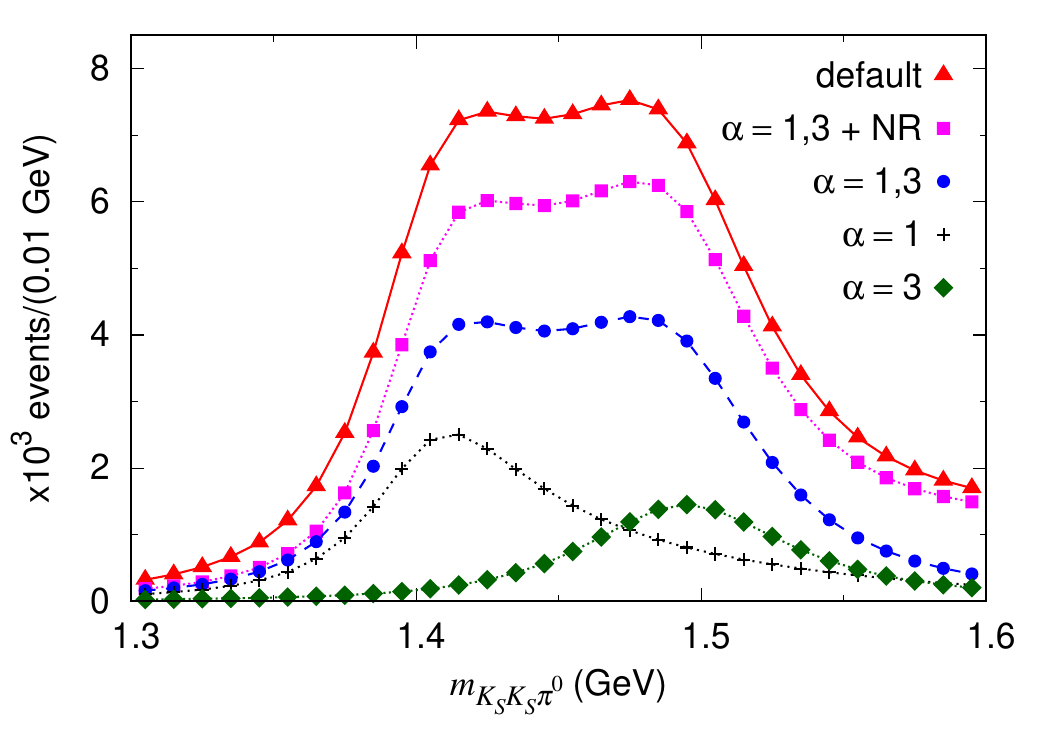}
\end{center}
 \caption{The pole contributions from $\eta(1405)$ ($\alpha$ = 1) and 
$\eta(1475)$ ($\alpha=3$) and their coherent sum 
($\alpha=1,3$) to
the $m_{K_SK_S\pi^0}$ distribution.
The poles labeled by $\alpha$ are listed in Table~\ref{tab:pole}.
The nonresonant (NR) contribution is from Eq.~(\ref{eq:cnr}).
 }
\label{fig:kkpi-pole}
\end{figure}
We examine the resonance pole contributions to 
the $E$ distribution.
For this purpose, we expand the dressed 
$\eta^*$ propagator of Eq.~(\ref{eq:mstar-g1}) around the resonance pole
at $M_{R_\alpha}$ as~\cite{SSL2},
\begin{eqnarray}
\label{eq:res-expand}
\left[\bar{G}(E)\right]_{ij} &\sim& {\chi_{\alpha,i}\,\chi_{\alpha,j} \over E - M_{R_\alpha} } , 
\end{eqnarray}
with
\begin{eqnarray}
\chi_{\alpha,1} &=& \sqrt{M_{R_\alpha} - m_{\eta^*_2} - [\Sigma_{\eta^*}(M_{R_\alpha})]_{22}\over \Delta'(M_{R_\alpha})}, \\
\chi_{\alpha,2} &=& {[\Sigma_{\eta^*}(M_{R_\alpha})]_{12}\over 
M_{R_\alpha} - m_{\eta^*_2} - [\Sigma_{\eta^*}(M_{R_\alpha})]_{22}}\chi_{\alpha,1}, 
\label{eq:mstar-expand}
\end{eqnarray}
$\Delta(E) \equiv {\rm det}[\bar{G}^{-1}(E)]$, and 
$\Delta'(M_{R_\alpha}) = d \Delta(E)/dE|_{E=M_{R_\alpha}}$.
Then we replace 
$\bar{G}_{ij}(E)$ in the full amplitude of Eq.~(\ref{eq:amp_a}) with
the above expanded form, and calculate the 
 $m_{K_SK_S\pi^0}$ distribution.
In Fig.~\ref{fig:kkpi-pole}, we show
each of the pole contributions and their
coherent sum, in comparison with the full calculation.
The $\alpha=2$ pole contribution is not included in the figure
since the $K^*\bar{K}$ branch cut mostly screens this pole contribution to
the amplitude on the physical real $E$ axis.
The contributions from the $\alpha=1$ and 3 poles are dominant, and the lineshape of the full
calculation is mostly formed by the the pole contributions.
The nonresonant term in Eq.~(\ref{eq:cnr}) enhances the spectrum overall
through the interference. 
Still, the branch cuts and non-pole contribution 
are missing in the pole approximation of 
Eq.~(\ref{eq:res-expand}),
and their effects should explain
the difference between the red triangles and the magenta squares in the figure.

\begin{figure}[t]
\begin{center}
\includegraphics[width=.5\textwidth]{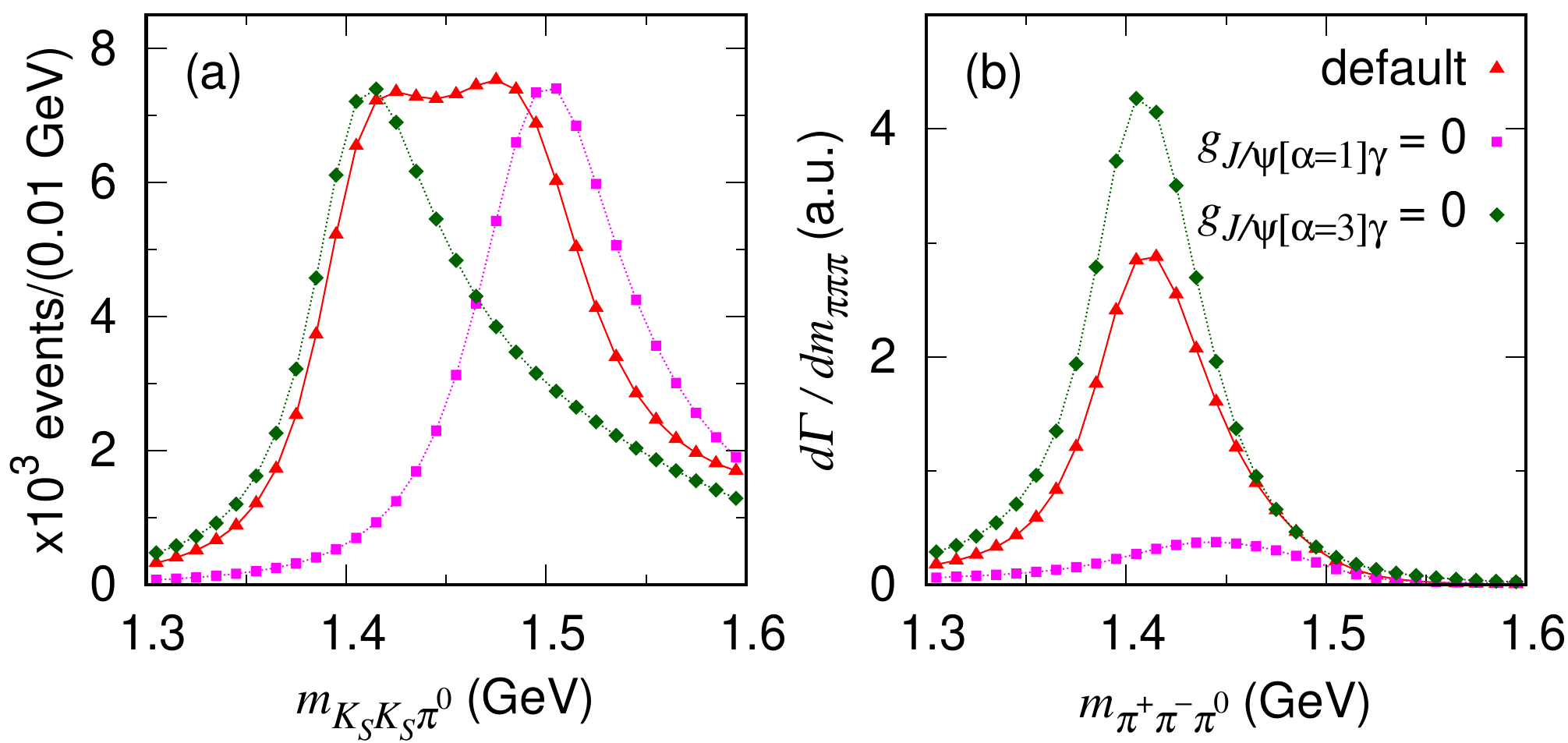}
\end{center}
 \caption{(a) $m_{K_SK_S\pi^0}$ and (b) $m_{\pi^+\pi^-\pi^0}$ distributions
from $J/\psi\to\gamma (0^{-+})\to\gamma(K_SK_S\pi^0)$ and 
$\gamma(\pi^+\pi^-\pi^0)$, respectively, 
obtained with various choices of 
$g_{J/\psi\eta^*_j\gamma}$ in Eq.~(\ref{eq:invamp2}).
The red triangles in (a) and (b) are the same as those in 
Fig.~\ref{fig:kkpi}(a) and \ref{fig:3pix}(a), respectively.
With the coupling to the $\alpha=1$ [$\alpha=3$] pole
eliminated, the magenta squares [green diamonds] are obtained..
All the calculations in (a)
use normalized $g_{J/\psi\eta^*_j\gamma}$ values
to have the same peak height.
The same legends in (a) and (b) share the same $g_{J/\psi\eta^*_j\gamma}$.
 }
\label{fig:pole-adjusted}
\end{figure}
The resonance amplitude of Eq.~(\ref{eq:res-expand}) suggests that 
one of the pole contributions 
can be eliminated 
from our full model
by adjusting the
coupling $g_{J/\psi\eta^*_j\gamma}$ in the 
initial vertex of Eq.~(\ref{eq:invamp2}).
Specifically, we can 
eliminate the contribution of the pole $\alpha$ by setting
\begin{eqnarray}
\label{eq:init_adj}
g_{J/\psi\eta^*_2\gamma}= -(\chi_{\alpha,1}/\chi_{\alpha,2})
g_{J/\psi\eta^*_1\gamma},
\end{eqnarray}
as demonstrated in Fig.~\ref{fig:pole-adjusted}(a).
The figure shows a full calculation without the pole
approximation of Eq.~(\ref{eq:res-expand}).
Eliminating the initial radiative transition of $J/\psi\to [\alpha=1]$,
we obtain the magenta squares ($g_{J/\psi[\alpha=1]\gamma}=0$)
showing a single peak from the 
$\alpha=3$ pole.
Similarly, a calculation with
$g_{J/\psi[\alpha=3]\gamma}=0$ gives
the green diamonds that have
a single peak from the $\alpha=1$ pole.

Among various processes that include
$\eta(1405/1475)$-decay into $K\bar{K}\pi$ final states,
some of them show a single peak from 
either of $\eta(1405)$ or $\eta(1475)$,
and others have a broad peak from a coherent sum of them.
Figure~\ref{fig:pole-adjusted}(a)
indicates that our coupled-channel model
can describe both cases 
by appropriately adjusting the couplings of 
initial vertices.

\begin{figure*}
\begin{center}
\includegraphics[width=.75\textwidth]{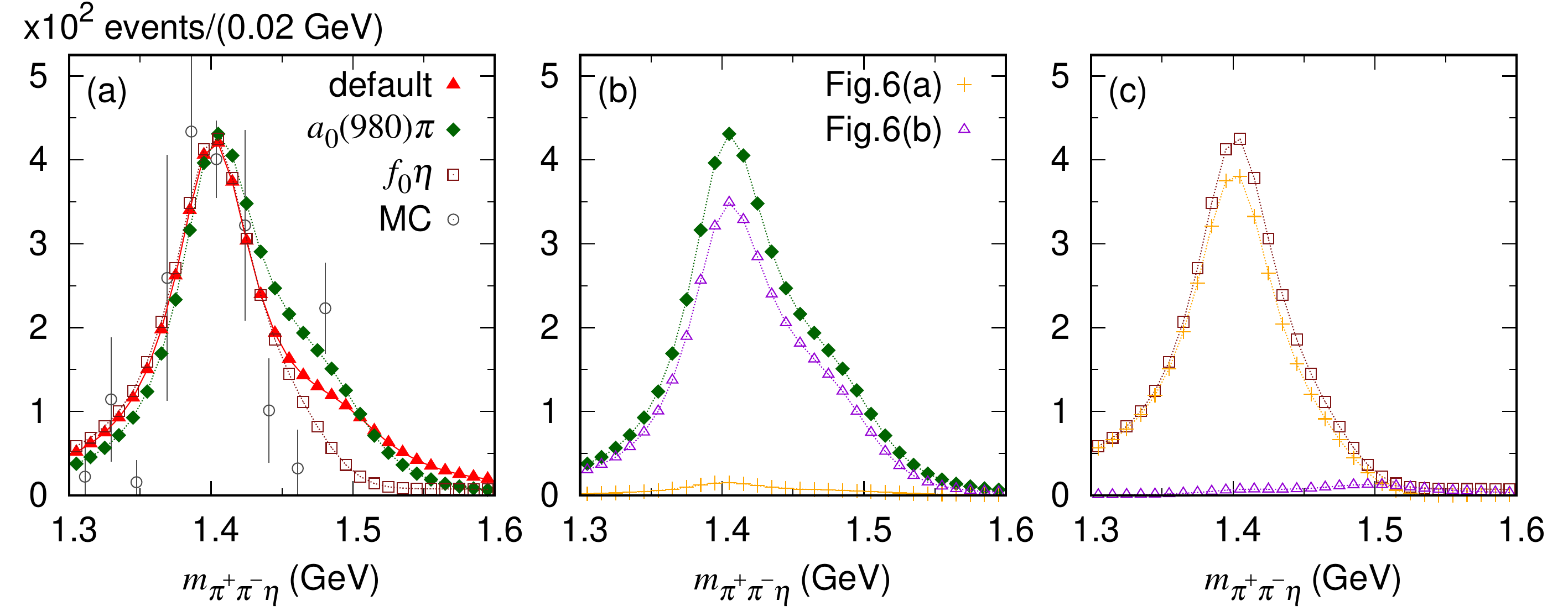}
\end{center}
 \caption{
The $m_{\pi\pi\eta}(=E)$ distribution for 
$J/\psi\to\gamma (0^{-+})\to\gamma(\pi^+\pi^-\eta)$.
(a) The default model prediction and
final $a_0(980)\pi$ and $f_0\eta$ contributions are shown.
The MC output is from Ref.~\cite{mark3_jpsi-gamma-eta-pipi}.
(b)[(c)] The final $a_0(980)\pi$ [$f_0\eta$] contribution.
Main contributions from the diagrams of Figs.~\ref{fig:diag2}(a) and
 \ref{fig:diag2}(b) are also shown.
Figure~\ref{fig:pipieta}(a) is taken from Ref.~\cite{letter}. Copyright (2023) APS.
}
\label{fig:pipieta}
\end{figure*}

In the presented analysis,
two bare states are required 
for reasonably fitting the dataset.
The lighter bare mass is determined to be $\sim 1.6$~GeV,
while the heavier one being $\sim 2.3$~GeV, as listed in Table~\ref{tab:param}
of Appendix. 
The heavier bare mass is not tightly constrained by the fit, and those
in the range of 2--2.4~GeV can give comparable fits.
Within our coupled-channel model, the bare states are mixed and dressed by meson-meson
continuum states, forming the resonance states.
In concept, the bare states are similar to 
states from a quark model or LQCD without two-hadron operators.
The lighter bare state seems compatible with the excited
$s\bar{s}$~\cite{pdg,barnes1997,dudek2013}.
The heavier bare state could be either of a second radial excitation of
$\eta^{(\prime)}$, a hybrid~\cite{dudek2013}, a
glueball~\cite{bali1993,morningstar1999,chen2006,richards2010,chen2111},
or a mixture of these states.

\begin{figure*}
\begin{center}
\includegraphics[width=.83\textwidth]{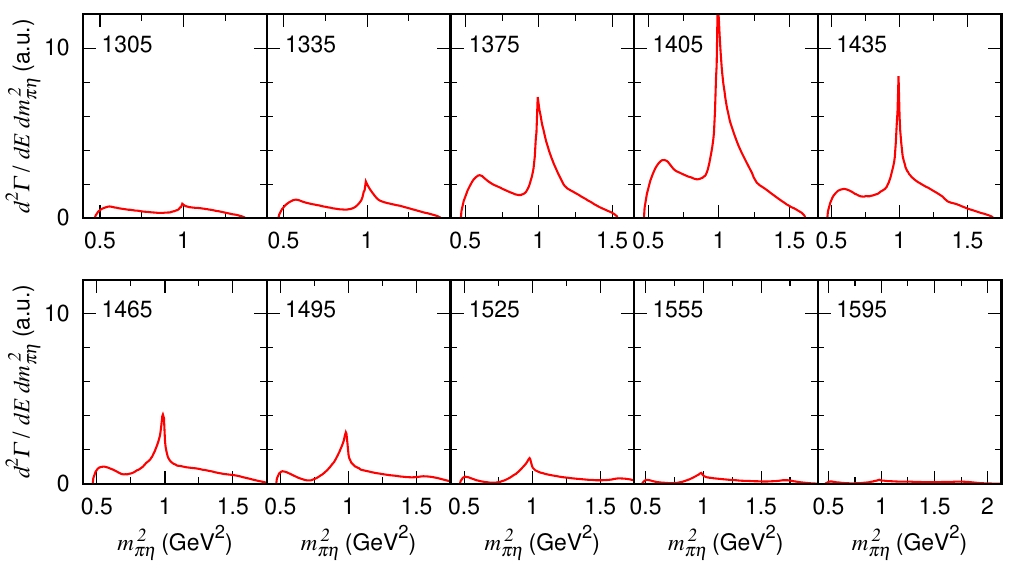}
\end{center}
 \caption{
The $m^2_{\pi\eta}$ distribution for 
$J/\psi\to\gamma(0^{-+})\to\gamma(\pi^+\pi^-\eta)$
from the default model.
The $E$ value is indicated in each panel.
}
\label{fig:pieta}
\end{figure*}

\section{Predictions for 
$J/\psi\to\gamma (0^{-+})\to\gamma(\pi\pi\eta), \gamma(\pi^+\pi^-\gamma), \gamma(\pi\pi\pi)$
}
\label{sec:prediction}

In this section, 
we present $E$ dependences of various final states from the radiative
$J/\psi$ decays via $\eta(1405/1475)$,
using the three-body unitary coupled-channel model 
developed in the previous section.
The model has been fitted to
the $K_SK_S\pi^0$ Dalitz plot pseudodata (Fig.~\ref{fig:fit1})
and the ratios of Eqs.~(\ref{eq:R1exp}) and (\ref{eq:R2exp}).

\subsection{$\pi^+\pi^-\eta$ and $\pi^0\pi^0\eta $ final states}

We show in Fig.~\ref{fig:pipieta}(a) the $m_{\pi\pi\eta}(=E)$ distributions for the
$\pi^+\pi^-\eta$ final state;
the $\pi^0\pi^0\eta$ distribution is smaller by a factor of 1/2.
The lineshape is qualitatively consistent with the MARK III
analysis~\cite{mark3_jpsi-gamma-eta-pipi}.
The final $a_0(980)\pi$ and $f_0\eta$ states have comparable
contributions. 
On the other hand, 
the $K\bar{K}\pi$ final state are mainly from 
the final $K^*\bar{K}$ and $\kappa\bar{K}$ contributions, as seen in Fig.~\ref{fig:kkpi}(a).
Since different $Rc$ final states couple with
$\eta(1405)$ and $\eta(1475)$ differently, 
the $K\bar{K}\pi$ and $\pi\pi\eta$ final states 
have different $E$ dependences.
The $\pi\pi\eta$ final states give a single peak at 
$m_{\pi\pi\eta}\sim$ 1.4~GeV, while
the $K\bar{K}\pi$ distribution has a flat peak. 
The process-dependent lineshape of 
the $\eta(1405/1475)$ decays can thus be understood.

In Figs.~\ref{fig:pipieta}(b) and \ref{fig:pipieta}(c), we decompose
the final $a_0(980)\pi$ and $f_0\eta$ contributions into
direct decays [Fig.~\ref{fig:diag2}(a)]
and single-rescattering mechanisms [Fig.~\ref{fig:diag2}(b)].
The final $a_0(980)\pi$ state is mostly from the single-rescattering mechanisms
and the direct decays are minor.
On the other hand, a completely opposite trend applies to the final $f_0\eta$ state.
In more detail, the $K^*K\bar{K}$, $\kappa K\bar{K}$, and 
$f_0\pi\eta$ triangle mechanisms contribute to the final 
$a_0(980)\pi$ state.
We find that the three loops give comparable contributions, 
even though only
the $K^*K\bar{K}$ loop causes a triangle singularity.
This is perhaps because the $K^*\bar{K}$ pair is relatively $p$-wave,
suppressing the triangle singularity. 

We also present in Fig.~\ref{fig:pieta}
a prediction for the $m^2_{\pi\eta}$ distribution 
from the default model. 
Clear $a_0(980)$ peaks are predicted, which is qualitatively consistent with 
the data~\cite{bes_jpsi-gamma-eta-pipi}.
This prediction should be confronted with the future data from the BESIII.

As already discussed, the final $a_0(980)\pi$ contribution to the $K\bar{K}\pi$
and $\pi\pi\eta$ final states are related
by the relative coupling of $a_0(980)\to K\bar{K}$ to 
$a_0(980)\to \pi\eta$ determined
experimentally~\cite{pdg,a0_980_ppbar,bes3_a0,omeg_a0,a0_980_gg}.
As we have seen in Fig.~\ref{fig:kkpi}(a),
the final $a_0(980)\pi$ contribution to $K\bar{K}\pi$
is very small to satisfy the ratio of Eq.~(\ref{eq:R1exp}).
If the final $a_0(980)\pi$ contribution to $K\bar{K}\pi$ were
as large as that of the BESIII amplitude model,
then Eq.~(\ref{eq:R1exp}) requires that 
the final $a_0(980)\pi\to\pi\pi\eta $ amplitude has to be 
drastically canceled by 
destructively interfering with the final $f_0\eta\to\pi\pi\eta$ amplitude. 
Such a large cancellation seems unlikely since there is no symmetry behind.
Also, the large cancellation makes the 
$a_0(980)$ peak in the $m_{\pi\eta}$ distribution rather unclear, but 
the data~\cite{bes_jpsi-gamma-eta-pipi} shows a clear $a_0(980)$ peak.
As shown in Fig.~\ref{fig:pieta},
our default model creates a clear $a_0(980)$ peak.

\subsection{$\pi^+\pi^-\gamma$ final state}

\begin{figure}
\begin{center}
\includegraphics[width=.49\textwidth]{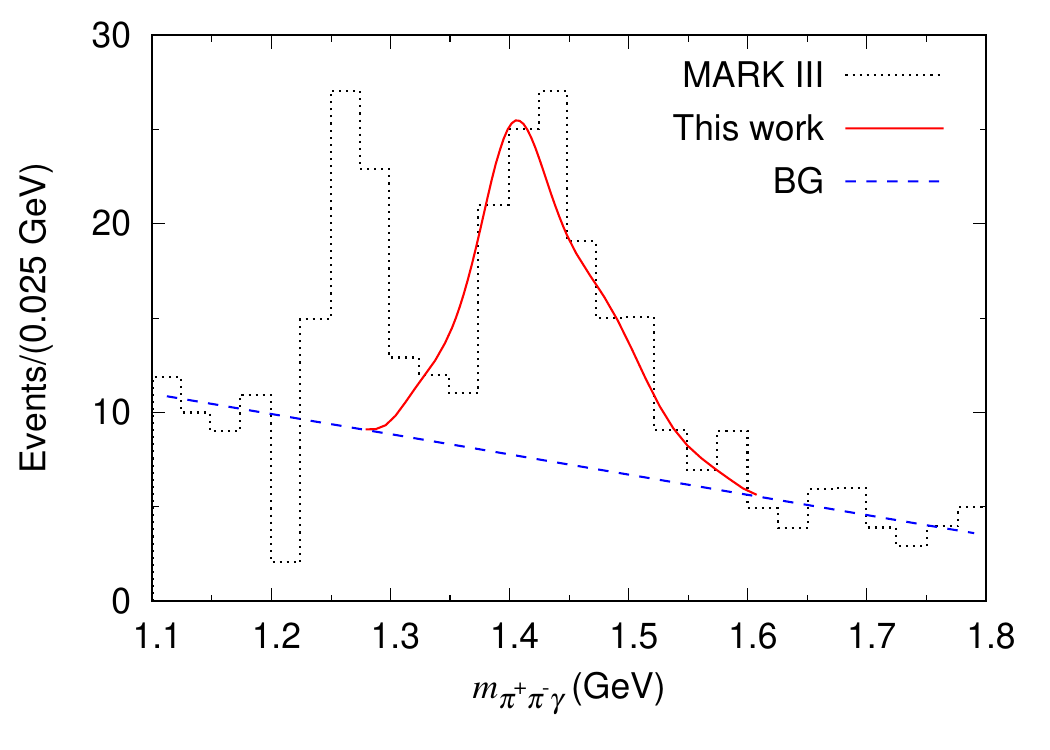}
\end{center}
 \caption{
The $m_{\pi^+\pi^-\gamma}(=E)$ distribution of 
$J/\psi\to\gamma (0^{-+})\to\gamma(\pi^+\pi^-\gamma)$ where 
the $\pi^+\pi^-$ pair is from $\rho^0$ decay.
The default model predicts the lineshape of the red curve which has been
 smeared with the experimental bin width, scaled by a factor, and
 augmented by a linear background (BG) to fit the data~\cite{mark3_rhog}.
}
\label{fig:rho_gamma}
\end{figure}

The branching to $J/\psi\to\gamma (0^{-+})\to\gamma(\pi^+\pi^-\gamma)$ 
in the default model
is constrained by the ratio of Eq.~(\ref{eq:R2exp}).
Then the model predicts the $E$ distribution 
as shown in Fig.~\ref{fig:rho_gamma}.
The lineshape has a single peak at $E\sim 1.4$~GeV, 
being consistent with the previous data~\cite{mark3_rhog,bes2_rhog}.
The process is mostly from a sequence of 
$\eta^*_i\to\rho^0\rho^0$
followed by 
$\rho^0\to\gamma$ and $\rho^0\to\pi^+\pi^-$.
Thus $\eta(1405)$ couples to $\rho\rho$
much more strongly than 
$\eta(1475)$ does,
implying different natures of the two $\eta^*$ resonances.
Also, as mentioned in Sec.~\ref{subsec:model},
only the heavier bare $\eta^*$ couples with $\rho\rho$.
This implies that 
$\eta(1405)$ includes a larger content of 
the heavier bare $\eta^*$ than 
$\eta(1475)$ does.

\subsection{$\pi^+\pi^-\pi^0$ and $\pi^0\pi^0\pi^0$ final states}

\begin{figure*}
\begin{center}
\includegraphics[width=1\textwidth]{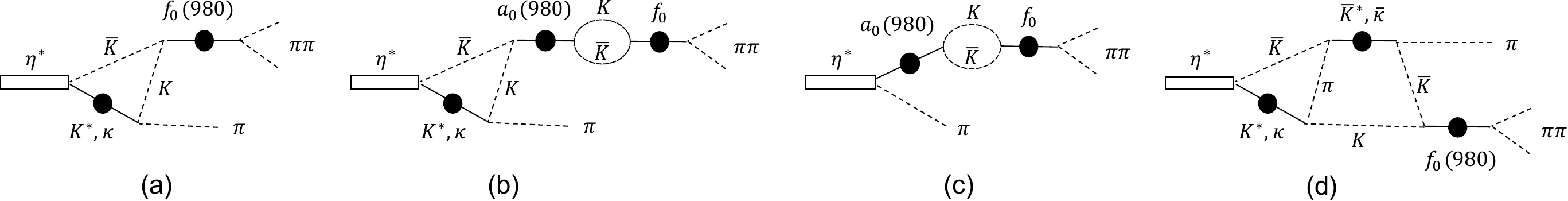}
\end{center}
 \caption{Main mechanisms for isospin-violating $\eta^*\to\pi\pi\pi$ decay included in
 Fig.~\ref{fig:diag}(c):
(a) isospin-violating $K^*(\kappa)\bar{K}K$ triangle loop;
(b) isospin-conserving $K^*(\kappa)\bar{K}K$ triangle loop followed by $a_0$-$f_0$ mixing;
(c) direct decay to $a_0(980)\pi$ followed by $a_0$-$f_0$ mixing;
(d) isospin-conserving $K^*(\kappa)\bar{K}\pi$ triangle loop followed by
isospin-violating $\pi\bar{K}^*(\bar{\kappa})K\bar{K}$ box loop.
 }
\label{fig:diag-3pi}
\end{figure*}

\begin{figure}
\begin{center}
\includegraphics[width=.49\textwidth]{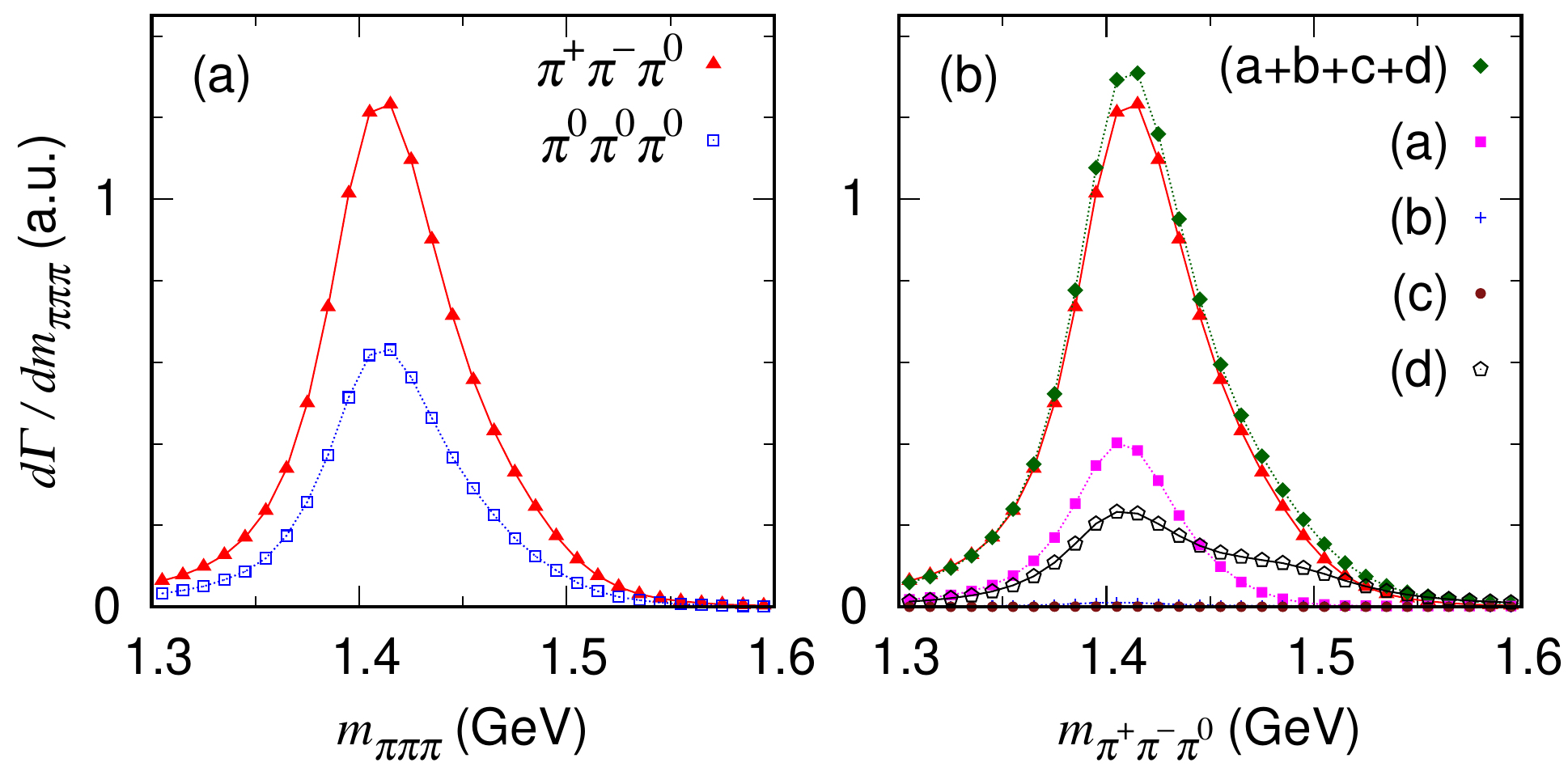}
\end{center}
 \caption{The $m_{\pi\pi\pi}(=E)$ distributions for 
$J/\psi\to \gamma (0^{-+})\to\gamma(\pi\pi\pi)$
predicted by the default model.
(a) The lineshapes and relative magnitudes of the
$\pi^+\pi^-\pi^0$ and 
$\pi^0\pi^0\pi^0$ final states.
(b) Contributions to the $\pi^+\pi^-\pi^0$ final state
from the diagrams in Figs.~\ref{fig:diag-3pi}(a)-(d).
 }
\label{fig:3pix}
\end{figure}
Our default model makes predictions for
the isospin-violating $J/\psi\to\gamma (0^{-+})\to\gamma(\pi\pi\pi)$;
the model has not been constrained by any data of the 
$\pi\pi\pi$ final states.
These isospin-violating processes are mainly from the mechanisms
of Fig.~\ref{fig:diag-3pi} that are not completely canceled due to
the small difference between the charged and neutral $K$ masses.
In particular, the isospin-violating mechanisms in 
Figs.~\ref{fig:diag-3pi}(b) and \ref{fig:diag-3pi}(c) are called the
$a_0$-$f_0$ mixing.
The $m_{\pi\pi\pi}$ distributions are shown in Fig.~\ref{fig:3pix}(a).
The $\pi^+\pi^-\pi^0$ distribution 
is almost twice as large as 
the $\pi^0\pi^0\pi^0$ distribution.
The $m_{\pi\pi\pi}$ distributions have a single peak at $\sim 1.4$~GeV.

\begin{figure}[b]
\begin{center}
\includegraphics[width=.49\textwidth]{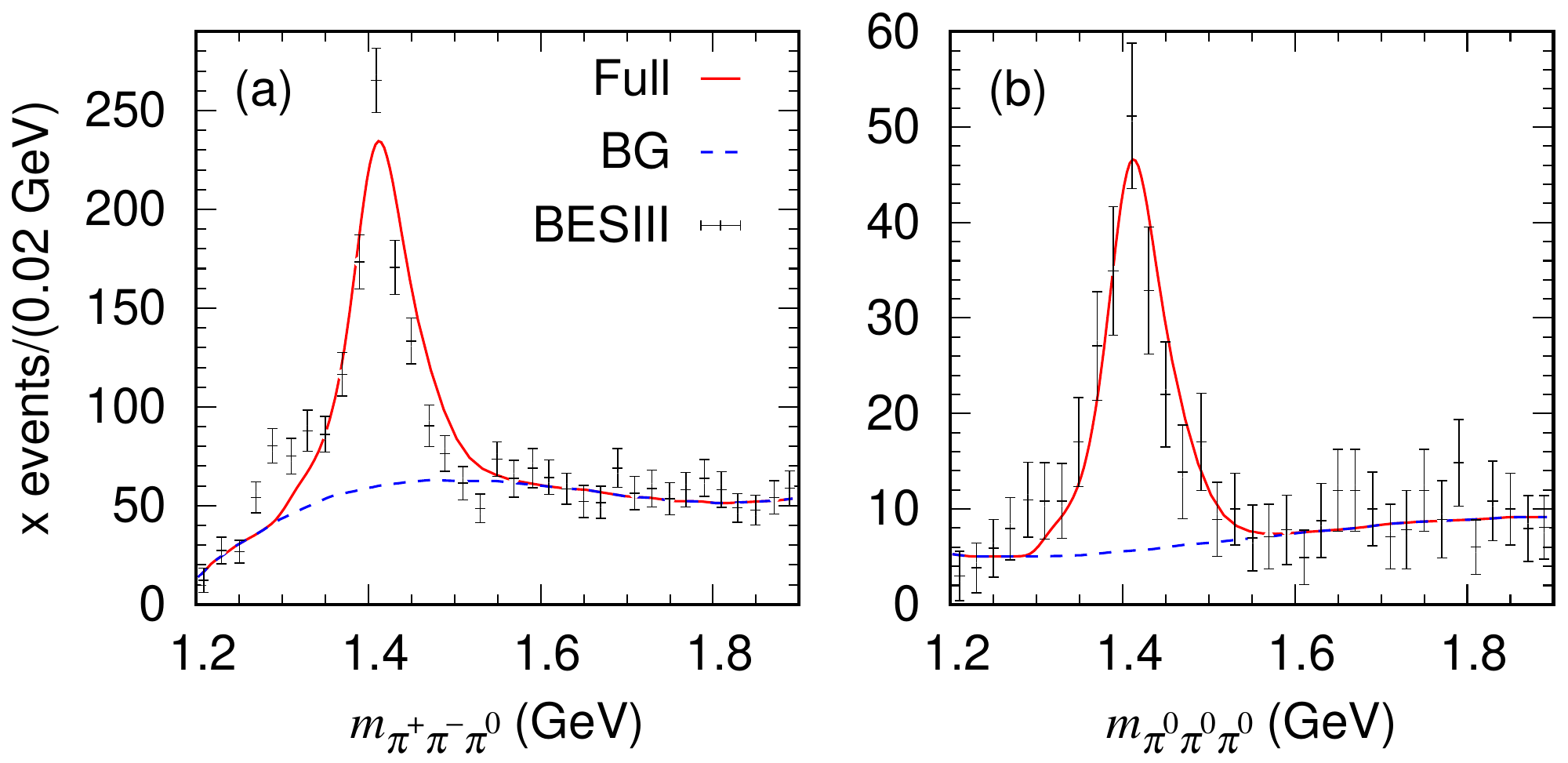}
\end{center}
 \caption{
The $m_{\pi\pi\pi}(=E)$ distributions for 
$J/\psi\to \gamma (0^{-+})\to\gamma(\pi\pi\pi)$
in comparison with the BESIII data~\cite{bes3-3pi} for
(a) $\pi^+\pi^-\pi^0$ and (b) $\pi^0\pi^0\pi^0$ final states. 
The full calculations are smeared with the bin width,
scaled to fit the data, and augmented by the background polynomials (BG)
from Ref.~\cite{bes3-3pi}.
 }
\label{fig:3pi}
\end{figure}

Contributions from the diagrams of Figs.~\ref{fig:diag-3pi}(a)-(d)
are separately shown in Fig.~\ref{fig:3pix}(b).
The $K^*(\kappa)\bar{K}K$ triangle loop
diagram of Fig.~\ref{fig:diag-3pi}(a) generates a clear peak.
As has been discussed in the literature,
this $K^*\bar{K}K$ triangle loop is
significantly enhanced by a TS occurring at $E\sim 1.40$~GeV.
The $\kappa\bar{K}K$ triangle loop without TS gives a smaller contribution.
The TS-enhancement is larger 
around the higher end of the TS energy range since 
the $p$-wave $K^*\bar{K}$ pair suppresses
the TS-enhancement around the lower end.
This explains the peak position in Fig.~\ref{fig:3pix}.

The $a_0$-$f_0$ mixing contribution is very small. 
This is because 
$\eta(1405/1475)\to a_0(980)\pi$ is very little as seen in Fig.~\ref{fig:kkpi}(a).
This small branching is required by the experimental ratio of 
Eq.~(\ref{eq:R1exp}).
The two-loop mechanisms of Fig.~\ref{fig:diag-3pi}(d) are sizable;
the second loop involves a TS.
A part of the two-loop contribution
is from mechanisms where the two loops are 
mediated by $v^{{\rm HLS}}$ in  Eq.~(\ref{eq:vvv}).
The coherent sum of the mechanisms in Fig.~\ref{fig:diag-3pi}
(green diamonds in Fig.~\ref{fig:3pix})
mostly explain the full calculation (red triangles).

We confront our predictions for the 
$\pi^+\pi^-\pi^0$ and $\pi^0\pi^0\pi^0$ lineshapes
with the BESIII data~\cite{bes3-3pi} in 
Figs.~\ref{fig:3pi}(a) and \ref{fig:3pi}(b), respectively.
Our model correctly predicts the peak position. 
This remarkable agreement suggests that the peak position is
determined by a kinematical effect (triangle singularity) that does
not depend on dynamical details.
However, the peak width from our calculation seems somewhat broader than the
data; we will come back to this point later. 

 \begin{figure}[b]
\begin{center}
\includegraphics[width=.49\textwidth]{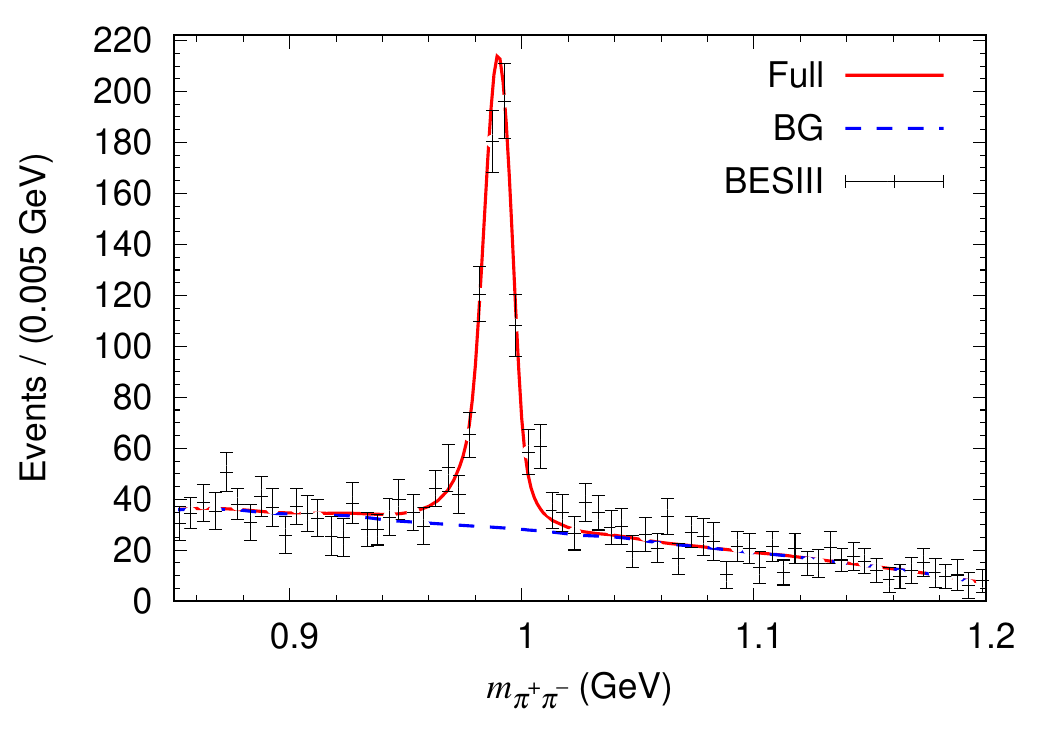}
\end{center}
 \caption{The $m_{\pi^+\pi^-}$ distribution of 
$J/\psi\to\gamma (0^{-+})\to\gamma(\pi^+\pi^-\pi^0)$.
The BESIII data and the background polynomial (BG)
are from Ref.~\cite{bes3-3pi}.
Our full calculation has been smeared with the bin width,
scaled to fit the data, and augmented by the background.
 }
\label{fig:3pi-5}
 \end{figure}
In Fig.~\ref{fig:3pi-5},
we also compare the $m_{\pi^+\pi^-}$ distribution
from our full calculation with the BESIII
data~\cite{bes3-3pi}.
Again, the agreement is reasonable, 
showing the sound predictive power of the coupled-channel model
that appropriately account for the relevant kinematical effect for the
isospin violation. 
The $f_0(980)$-like peak width ($\sim 10$~MeV) is much narrower than the
world average ($\sim 50$~MeV)~\cite{pdg}.
This occurs because 
the $(K^*)K^+K^-$ and $(K^*)K^0\bar{K}^0$ loops 
in Fig.~\ref{fig:diag-3pi} almost exactly cancel with
each other due to the isospin symmetry, except 
in a small window ($\sim$8~MeV) of 
$2 m_{K^\pm}< m_{\pi\pi} < 2 m_{K^0}$
where 
the two loops are rather different and the cancellation is
incomplete.
Furthermore, the TS enhances
the $f_0(980)$-like peak.
Therefore, 
the $f_0(980)$ pole plays a minor role in developing the peak
in Fig.~\ref{fig:3pi-5}.

How do $\eta(1405)$ and $\eta(1475)$ resonances work in 
$J/\psi\to\gamma(\pi\pi\pi)$ ?
We address this question by using the models shown in 
Fig.~\ref{fig:pole-adjusted}(a).
In the figure, the models labeled by 
$g_{J/\psi[\alpha=1]\gamma}=0$ and $g_{J/\psi[\alpha=3]\gamma}=0$ do not
have $J/\psi\to\gamma\eta(1405)$ and $J/\psi\to\gamma\eta(1475)$
couplings, respectively,
and they are normalized to have the same peak height in the
$m_{K_SK_S\pi^0}$ distribution.
Then, we use them to calculate
$J/\psi\to\gamma(\pi^+\pi^-\pi^0)$ as shown in
Fig.~\ref{fig:pole-adjusted}(b).
For the model of $g_{J/\psi[\alpha=3]\gamma}=0$,
the peak positions are almost the same for $K\bar{K}\pi$ and $\pi\pi\pi$
final states. 
This is because $\eta(1405)\to\pi\pi\pi$ is dominant and 
the $\eta(1405)$ mass and the TS region overlap well.
However, the peak width is narrower for $\pi\pi\pi$ because 
the TS region is narrower than the $\eta(1405)$ width.
On the other hand, the model of $g_{J/\psi[\alpha=1]\gamma}=0$ gives 
a significantly suppressed $m_{\pi\pi\pi}$ distribution in comparison with
the model of $g_{J/\psi[\alpha=3]\gamma}=0$.
This is because 
the $\eta(1475)$ mass is outside of the TS region and
$\eta(1475)\to\pi\pi\pi$ is not enhanced.
In this way, we understand how the different 
$K\bar{K}\pi$ and $\pi\pi\pi$ lineshapes in Fig.~\ref{fig:pole-adjusted}
are caused.

Finally, we compare ratios of $K\bar{K}\pi$ and $\pi\pi\pi$ branching fractions from our model with the
experimental counterpart. 
Using the $K\bar{K}\pi$ and $\pi\pi\pi$ branching ratios in Refs.~\cite{pdg,bes3-3pi},
we have experimental ratios:
\begin{eqnarray}
R_3^{\rm exp}
&&={
\Gamma  [J/\psi\to\gamma\eta(1405/1475)\to\gamma(\pi^+\pi^-\pi^0)]
\over 
\Gamma  [J/\psi\to\gamma\eta(1405/1475)\to\gamma(K\bar{K}\pi)]
}
\nonumber\\
&&=
{ 
(1.50\pm 0.11\pm 0.11)\times 10^{-5}
\over
(2.8\pm 0.6)\times 10^{-3} 
}
\nonumber\\
&&= 0.004-0.007 \ ,
\label{eq:R3exp}
\end{eqnarray}
and
\begin{eqnarray}
R_4^{\rm exp}
&&={
\Gamma  [J/\psi\to\gamma\eta(1405/1475)\to\gamma(\pi^0\pi^0\pi^0)]
\over 
\Gamma  [J/\psi\to\gamma\eta(1405/1475)\to\gamma(K\bar{K}\pi)]
}
\nonumber\\
&&= 
{ 
(7.10\pm 0.82\pm 0.72)\times 10^{-6}
\over
(2.8\pm 0.6)\times 10^{-3} 
}
\nonumber\\
&&= 0.002-0.003 \ .
\label{eq:R4exp}
\end{eqnarray}
Our coupled-channel model predicts:
\begin{eqnarray}
R_3^{\rm th} = 0.0020-0.0021, \qquad 
R_4^{\rm th} = 0.0010-0.0011, \qquad 
\label{eq:R4th}
\end{eqnarray}
which is significantly smaller than the data. 
A possible reason for the deficit is that 
we do not consider a contribution from the $J^{PC}=1^{++}$ partial
wave that includes $f_1(1285)$ and $f_1(1420)$.
The BESIII analysis~\cite{bes3_mc} found 
that 20-30\% of $J/\psi\to\gamma(K_SK_S\pi^0)$ is from 
the $1^{++}$ contribution in which 
$f_1(1420)\to K^*\bar{K}$ is a dominant mechanism.
Considering the consistency with
$J/\psi\to\gamma(K\bar{K}\pi)$,
$J/\psi\to\gamma(\pi\pi\pi)$ should come not only from the mechanisms of 
Fig.~\ref{fig:diag-3pi} but also from similar mechanisms that originate
from $f_1$ decays.
In particular, the triangle diagram from the $f_1(1420)$ decay
similar to Fig.~\ref{fig:diag-3pi}(a) 
would be significantly enhanced by the TS, 
since the $f_1(1420)$ mass and width have a good overlap with
the TS region. 
Furthermore, $f_1(1420)$ creates an $s$-wave $K^*\bar{K}$ pair 
while $\eta(1405)$ creates a $p$-wave pair. 
Thus the triangle mechanism from $f_1(1420)$
is more enhanced by the TS than that from $\eta(1405)$.
This $1^{++}$ contribution might explain the difference between our
prediction of Eq.~(\ref{eq:R4th}) and the experimental ratios of 
Eqs.~(\ref{eq:R3exp}) and (\ref{eq:R4exp}).
We also note that the BESIII~\cite{bes3-3pi} 
did not separate out
a possible $f_1(1420)$ contribution from
$\Gamma  [J/\psi\to\gamma\eta(1405/1475)\to\gamma(\pi\pi\pi)]$
in Eqs.~(\ref{eq:R3exp}) and (\ref{eq:R4exp}). 
The stronger TS enhancement would create a sharper peak in 
the $m_{\pi\pi\pi}$ lineshape.
In Fig.~\ref{fig:3pi},
our $0^{-+}$ model shows a peak somewhat broader than the data. 
By adding a sharper $1^{++}$ peak,
the data might be better fitted.

\section{Summary and future prospects}
\label{sec:summary}

Whether $\eta(1405/1475)$ is one or two states has been a controversial
issue. 
The recent BESIII amplitude analysis of $J/\psi\to\gamma K_SK_S\pi^0$ made important progress by claiming two states with a high confidence level. 
This analysis was based on $\sim 10^{10}$ $J/\psi$ decay samples which is significantly more precise than earlier $\eta(1405/1475)$-related data.
However, the BESIII analysis used a simple Breit-Wigner amplitude for $\eta(1405/1475)$.
For a more reliable determination of the $\eta(1405/1475)$ poles and their decay dynamics, 
three-body unitary coupled-channel analysis is desirable.

Thus, we developed a model for radiative $J/\psi$ decays to three pseudoscalar-meson final states of any partial wave ($J^{PC}$).
Also, a slight extension was made to include $\gamma\rho (\rho\to\pi^+\pi^-)$ final state. 
The main components of the model are two-body $\pi K$, $\pi\pi$, $K\bar{K}$, and $\pi\eta$ scattering models that generate $K^*_0(700)(=\kappa)$, $K^*(892)$, $f_0(500)(=\sigma)$, $f_0(980)$, $a_0(980)$, and $a_2(1320)$ resonance poles in the scattering amplitudes. 
The two-body scattering models as well as bare resonance states were implemented into the three-body coupled-channel scattering equation (Faddeev equation).
By solving the equation, we obtained the three-body unitary amplitudes with which we described the final-state interactions in the radiative $J/\psi$ decays. 

Using the BESIII's $J^{PC}=0^{-+}$ amplitude for $J/\psi\to\gamma K_SK_S\pi^0$, we generated $K_SK_S\pi^0$ Dalitz plot pseudo data for 30 energy bins in  1.3~GeV $\le m_{K_SK_S\pi^0}\le$ 1.6~GeV.
Then the pseudo data were fitted with the coupled-channel model.
The experimental branching ratios of $\eta(1405/1475)\to\eta\pi\pi$ and $\eta(1405/1475)\to\gamma\rho$ relative to that of $\eta(1405/1475)\to K\bar{K}\pi$ were simultaneously fitted. 
We obtained a reasonable fit with two bare $\eta^*$ states while, with
one bare $\eta^*$ state, we did not find a reasonable solution.
A noteworthy difference from the BESIII amplitude model is that the $a_0(980)\pi$ contribution is dominant (very small) in the BESIII (our) model.
The small $a_0(980)\pi$ contribution is required by the empirical branching ratio of $\eta(1405/1475)\to\eta\pi\pi$ that was not considered in the BESIII analysis.

Our $0^{-+}$ amplitude was analytically continued to reach three poles in the $\eta(1405/1475)$ region.
Two poles corresponding to $\eta(1405)$ were found near the $K^*\bar{K}$ threshold, and are located on different Riemann sheet of the $K^*\bar{K}$ channel. 
Another pole is $\eta(1475)$.
We made 50 bootstrap fits, and estimated statistical uncertainties of the pole positions (Table~\ref{tab:pole}).
This is the first pole determination of $\eta(1405/1475)$ and, furthermore, the first-ever pole determination from analyzing experimental Dalitz plot distributions with a manifestly three-body unitary coupled-channel framework.

The obtained model was used to predict the $\eta\pi\pi$ and $\gamma\pi^+\pi^-$ lineshapes of 
$J/\psi\to\gamma(0^{-+})\to \gamma(\eta\pi\pi)$ and $\gamma(\gamma\rho)$ processes.
The predicted lineshapes are process-dependent and reasonably consistent with the existing data.
We also applied the model to the isospin-violating $J/\psi\to\gamma(0^{-+})\to \gamma(\pi\pi\pi)$.
The importance of the triangle singularity from the $K^*\bar{K}K$ loop was clarified, while the $a_0(980)$-$f_0(980)$ mixing gave a tiny contribution.
Furthermore, the two-loop contribution was calculated for the first time,
and this contribution was shown to significantly enhance the isospin violation.
The predicted $\pi\pi\pi$ and $\pi^+\pi^-$ lineshapes agree well with the BESIII data.
Although the predicted branching fraction underestimates the data, we may expect the $1^{++}$ partial wave including $f_1(1420)$ to fill the deficiency.


Here, we stress that all of the above conclusions are based on the
Dalitz plot pseudo data including only the $0^{-+}$ contribution, and on the current
branching ratios of $\eta(1405/1475)\to\eta\pi\pi$ and
$\eta(1405/1475)\to\gamma\rho$
relative to that of $\eta(1405/1475)\to K\bar{K}\pi$. 
Since all of this experimental information was extracted with 
simpler Breit-Wigner models, our results might be biased.
This situation encourages further studies.


In the next step, we will extend the present analysis by including more partial waves such
as $1^{++}$ and $2^{++}$, and directly analyze the BESIII data of $J/\psi\to\gamma K_SK_S\pi^0$.
Then we can perform the partial wave decomposition with our
unitary coupled-channel framework by ourselves.
With the $0^{-+}$ amplitude obtained in this way, 
the two-pole solution of $\eta(1405/1475)$ needs to be reexamined.
Also, we can study the relevant resonances such as $\eta(1405/1475)$ and $f_1(1420)$ with the unitary coupled-channel framework consistently.

Our model can be easily applied to other decay processes that could
involve $\eta(1405/1475)$
by simply changing the initial vertex of Eq.~(\ref{eq:invamp2}) and
keeping the rest the same.
Those processes include
$\psi(2S)\to\omega(K\bar{K}\pi)$~\cite{omega_eta1405}, 
$\psi(2S)\to\phi(\eta\pi\pi)$~\cite{phi_eta1405}, 
$J/\psi\to\omega(\eta\pi^+\pi^-)$~\cite{omega_etapipi},
$J/\psi\to\omega(K\bar{K}\pi)$, $\phi (K\bar{K}\pi)$, $\eta (K_S^0K^\pm\pi^\mp)$~\cite{omega_phi_eta_eta1405},
and 
$\chi_{c0}\to\eta(\pi\pi\eta)$, $\eta (K\bar{K}\pi)$.
It would be important to analyze these various processes to establish the nature of $\eta(1405/1475)$.

\begin{acknowledgments}
We acknowledge Y. Cheng, M.-C. Du, S.-S. Fang, F.-K. Guo, Y.-P. Huang, H.-B. Li,
B. Liu, X.-R. Lyu, W.-B. Qian, L. Qiu, X.-Y. Shen, J.-J. Xie, G.-F. Xu,
Q. Zhao, and B.-S. Zou for useful discussions.
This work is in part supported by 
National Natural Science Foundation of China (NSFC) under contracts 
U2032103, 11625523, 12175239, 12221005, 12305087, and U2032111, 
and also by
National Key Research and Development Program of China under Contracts 2020YFA0406400,
Chinese Academy of Sciences under Grant No. YSBR-101 (J.J.W.), and
the Start-up Funds of Nanjing Normal University 
under Grant No. 184080H201B20 (Q.H.).
\end{acknowledgments}

\appendix

\section{Two-meson scattering models}\label{app1}
\subsection{Formulas}

\begin{table*}
\renewcommand{\arraystretch}{1.6}
\tabcolsep=4.mm
\caption{\label{tab:R} Description of two-meson scattering models.
Partial waves are specified by 
the orbital angular momentum $L$ and the isospin $I$.
}
\begin{tabular}{cccccc}
$R$  & $\{L,I\}$ &\# of bare states   & contact interaction & $R$-decay channels & \# of poles\\\hline
$f_0$      & $\{0,0\}$ & 2  & $\pi\pi,K\bar K\to \pi\pi,K\bar K$ & $\pi\pi$, $K\bar K$ & 3\\
$\rho(770)$& $\{1,1\}$ & 1  & - & $\pi\pi$ & 1\\
$\kappa$ ($K^*_0(700)$)& $\{0,1/2\}$  & 1  & $K\pi\to K\pi$ & $K\pi$& 1\\
-- & $\{0,3/2\}$  & 0  & $K\pi\to K\pi$ & - & 0\\
$K^*(892)$ & $\{1,1/2\}$ & 1  & $K\pi\to K\pi$ & $K\pi$ & 1\\
$a_0(980)$ & $\{0,1\}$ & 1  & - & $\eta\pi$, $K\bar K$ & 1\\
$a_2(1320)$& $\{2,1\}$ & 1  & - & $\eta\pi$, $K\bar K$, $\rho(770)\pi$ & 1\\
\end{tabular}
\end{table*}

We develop a unitary coupled-channel model for each of 
$\pi\pi$, $\pi K$, and $\pi\eta$ partial wave scatterings.
Let us consider a $ab\to a'b'$ scattering 
with total energy $E$. A partial wave is specified by
the total angular momentum $L$, total isospin $I$.
The incoming and outgoing momenta are denoted by $q$ and $q'$,
respectively.
Suppose that the scattering can be described with a contact interaction of:
\begin{eqnarray}
v^{LI}_{a'b',ab} (q',q) = 
w^{LI}_{a'b'}(q') h^{LI}_{a'b',ab}\; w^{LI}_{ab}(q) ,
\label{eq:cont-ptl}
\end{eqnarray}
where $h^{LI}_{a'b',ab}$ is a coupling constant.
We also introduced a vertex function $w^{LI}_{ab}(q)$ in the form of:
\begin{eqnarray}
w^{LI}_{ab}(q) = 
{1\over \sqrt{{\cal B}_{ab}}}
{ [1 +(q/b^{LI}_{ab})^2]^{-2-L/2} \over \sqrt{E_a(q)E_b(q)}} 
\left(q\over m_\pi\right)^L  ,
\label{eq:vf-cont}
\end{eqnarray}
with $b^{LI}_{ab}$ being a cutoff;
${\cal B}_{ab}$ is a factor associated with the Bose symmetry:
${\cal B}_{ab}=1/2$ for identical particles $a$ and $b$, and 
${\cal B}_{ab}=1$ otherwise.
The partial wave amplitude is then given by
\begin{eqnarray}
t^{LI}_{a'b',ab} (q',q; E) &=& \sum_{a''b''}
w^{LI}_{a'b'}(q') \tau^{LI}_{a'b',a''b''}(E)\;
\nonumber\\
&&\times h^{LI}_{a''b'',ab}\; w^{LI}_{ab}(q) ,
\label{eq:pw-2body-cont}
\end{eqnarray}
with
\begin{eqnarray}
\left[(\tau^{LI}(E))^{-1}\right]_{a'b',ab} &=& 
\delta_{a'b',ab} -  \sigma^{LI}_{a'b',ab}(E) ,
\end{eqnarray}
\begin{eqnarray}
\sigma^{LI}_{a'b',ab}(E) &=& 
\int  dq\; q^2  
{ {\cal B}_{ab}\, h^{LI}_{a'b',ab}\left[w^{LI}_{ab}(q)\right]^2
\over E-E_a(q)-E_b(q)+i\epsilon}
\ .
\label{eq:pw-cont-self}
\end{eqnarray}

Next, we also include bare $R$-excitation mechanisms in the interaction as
\begin{eqnarray}
V^{LI}_{a'b',ab} (q',q; E) &=&
 \sum_{R} f^{LI}_{a'b',R}(q') {1\over E-m_R} f^{LI}_{R,ab}(q) 
\nonumber\\
&&
+ v^{LI}_{a'b',ab} (q',q)  ,
\label{eq:pw-2body-v}
\end{eqnarray}
with $m_R$ being the bare $R$ mass.
A bare $R\to ab$ vertex function is denoted by
${f}^{LI}_{ab,R}(q)$ and
$f^{LI}_{R,ab}(q)= f^{{LI}}_{ab,R}(q)$;
an explicit form has been given in Eq.~(\ref{eq:pipi-vertex}).
With the interaction of Eq.~(\ref{eq:pw-2body-v}),
the resulting scattering amplitude is given by
\begin{eqnarray}
T^{LI}_{a'b',ab} (q',q; E) &=&
 \sum_{R',R}
\bar{f}^{LI}_{a'b',R'}(q';E) \tau^{LI}_{R',R}(0,E) \bar{f}^{LI}_{R,ab}(q;E) 
\nonumber\\
&&+t^{LI}_{a'b',ab} (q',q; E) 
\ .
\label{eq:pw-2body-t}
\end{eqnarray}
The second term has been given in Eq.~(\ref{eq:pw-2body-cont}).
The dressed $R\to ab$ vertex, denoted by $\bar{f}_{ab,R}$,
is given by
\begin{eqnarray}
\bar{f}^{LI}_{ab,R}(q;E) &=& 
f^{LI}_{ab,R}(q) 
+ \sum_{a'b'}\int dq' q'^2\; 
\nonumber\\
&& 
\times{{\cal B}_{a'b'}\,t^{LI}_{ab,a'b'} (q,q'; E)\; f^{LI}_{a'b',R}(q')
\over E-E_{a'}(q')-E_{b'}(q')+ i\epsilon}  \ ,\\
\bar{f}^{LI}_{R,ab}(q;E) &=& 
f^{LI}_{R,ab}(q) + \sum_{a'b'}
\int dq' q'^2\; 
\nonumber\\
&&\times
{{\cal B}_{a'b'}\,f^{LI}_{R,a'b'}(q')\; t^{LI}_{a'b',ab} (q',q; E)
\over E-E_{a'}(q')-E_{b'}(q')+ i\epsilon} 
\ .
\label{eq:dressed-vertex}
\end{eqnarray}
The dressed Green function for $R$, $\tau^{LI}_{R',R}(p,E)$, 
in Eq.~(\ref{eq:pw-2body-t})
has been given in Eqs.~(\ref{eq:green-Rc}) and (\ref{eq:RR-self}) 
with $f_{ab,R'}$ being replaced by $\bar{f}_{ab,R'}$.

The partial wave amplitude, $T^{LI}_{a'b',ab}$ in Eq.~(\ref{eq:pw-2body-t}),
 is related to the $S$-matrix by
\begin{eqnarray}
s^{LI}_{ab,ab} (E)
 &&= \eta_{LI}\, e^{2i\delta_{LI}} \nonumber\\
&&= 
1 - 2\pi i\rho_{ab}\, {\cal B}_{ab} T^{LI}_{ab,ab} (q_o,q_o; E) \ ,
\label{eq:s-matrix}
\end{eqnarray}
where $\delta_{LI}$ and $\eta_{LI}$
are the phase shift and inelasticity, respectively;
$q_o$ is the on-shell momentum ($E=E_a(q_o)+E_b(q_o)$);
$\rho_{ab}= q_o E_a(q_o)E_b(q_o)/E$ is the phase-space factor.

\subsection{Fits to $\pi\pi$, $\pi K$, and $\pi\eta$ scattering data}
\label{sec:app-pipi}

\begin{figure*}
\begin{center}
\includegraphics[width=.32\textwidth]{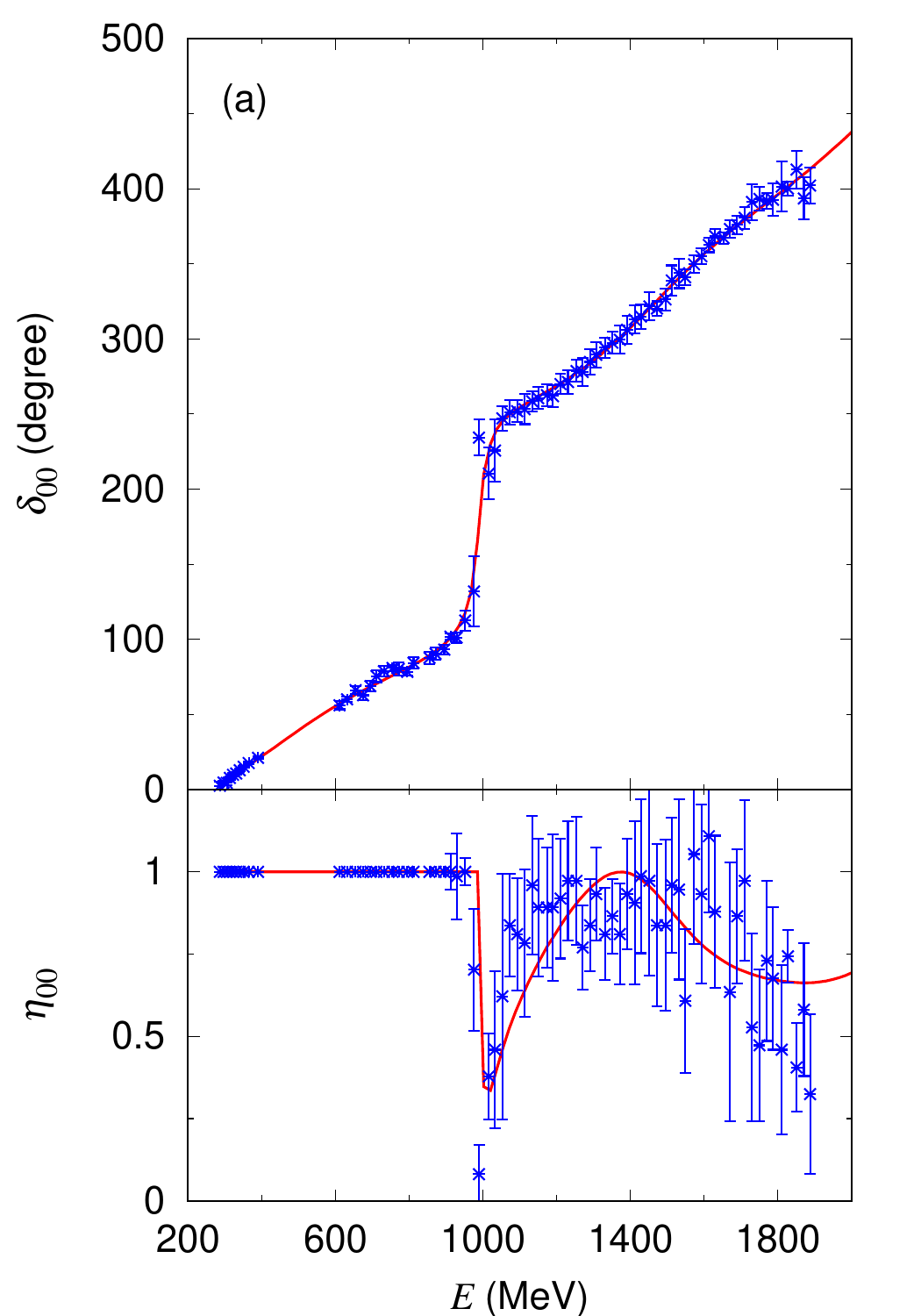}
\includegraphics[width=.32\textwidth]{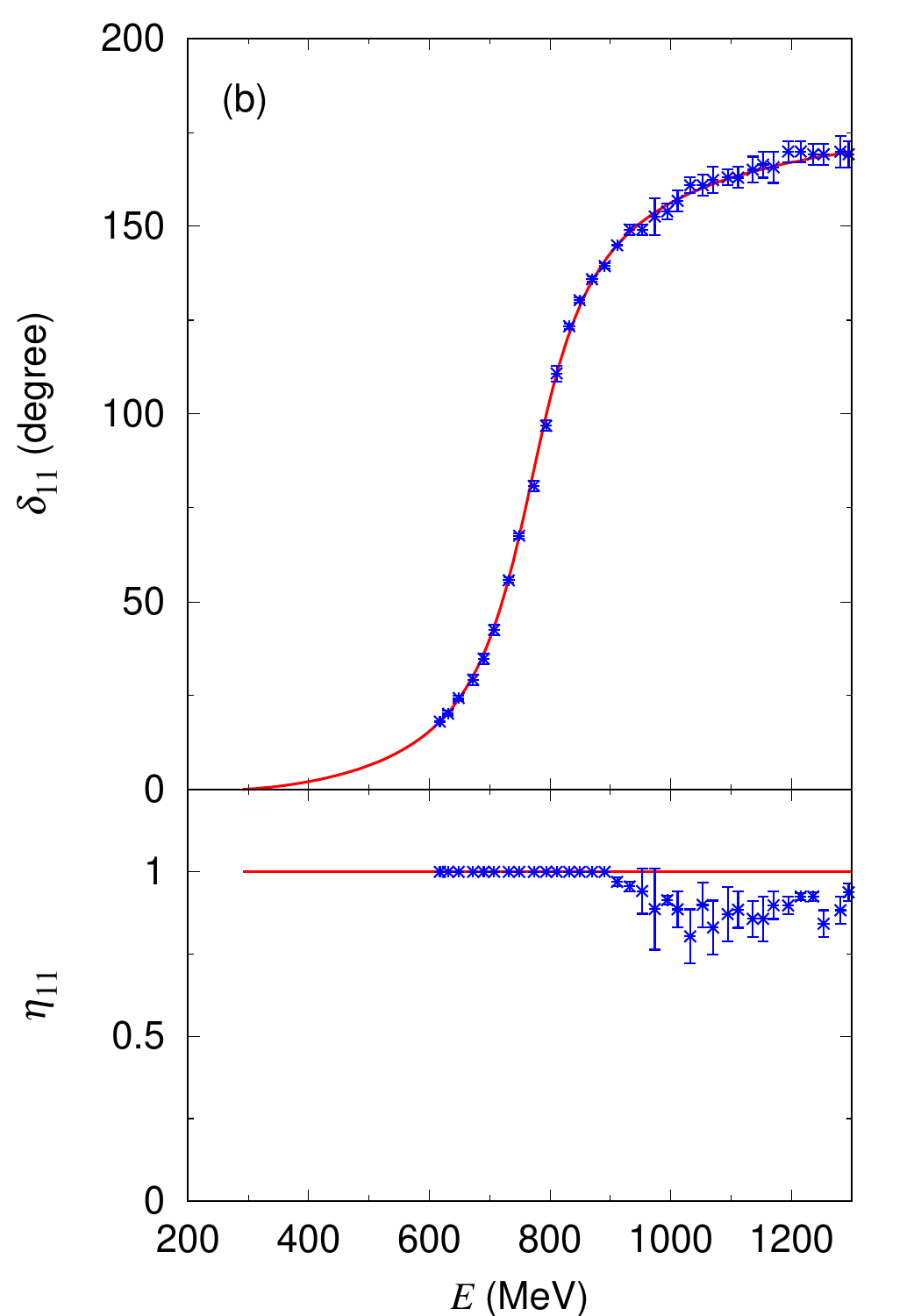}
\includegraphics[width=.32\textwidth]{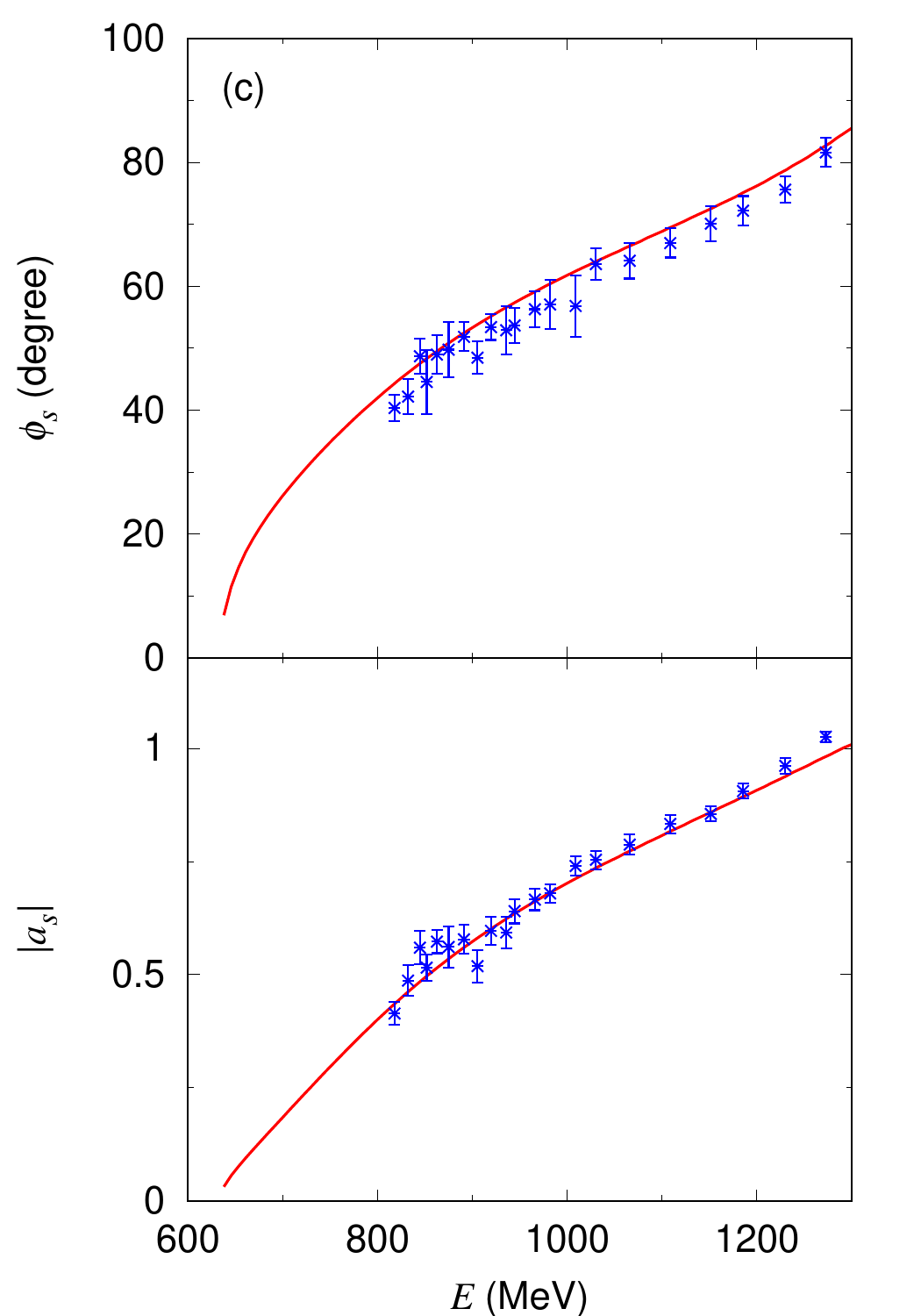}
\includegraphics[width=.32\textwidth]{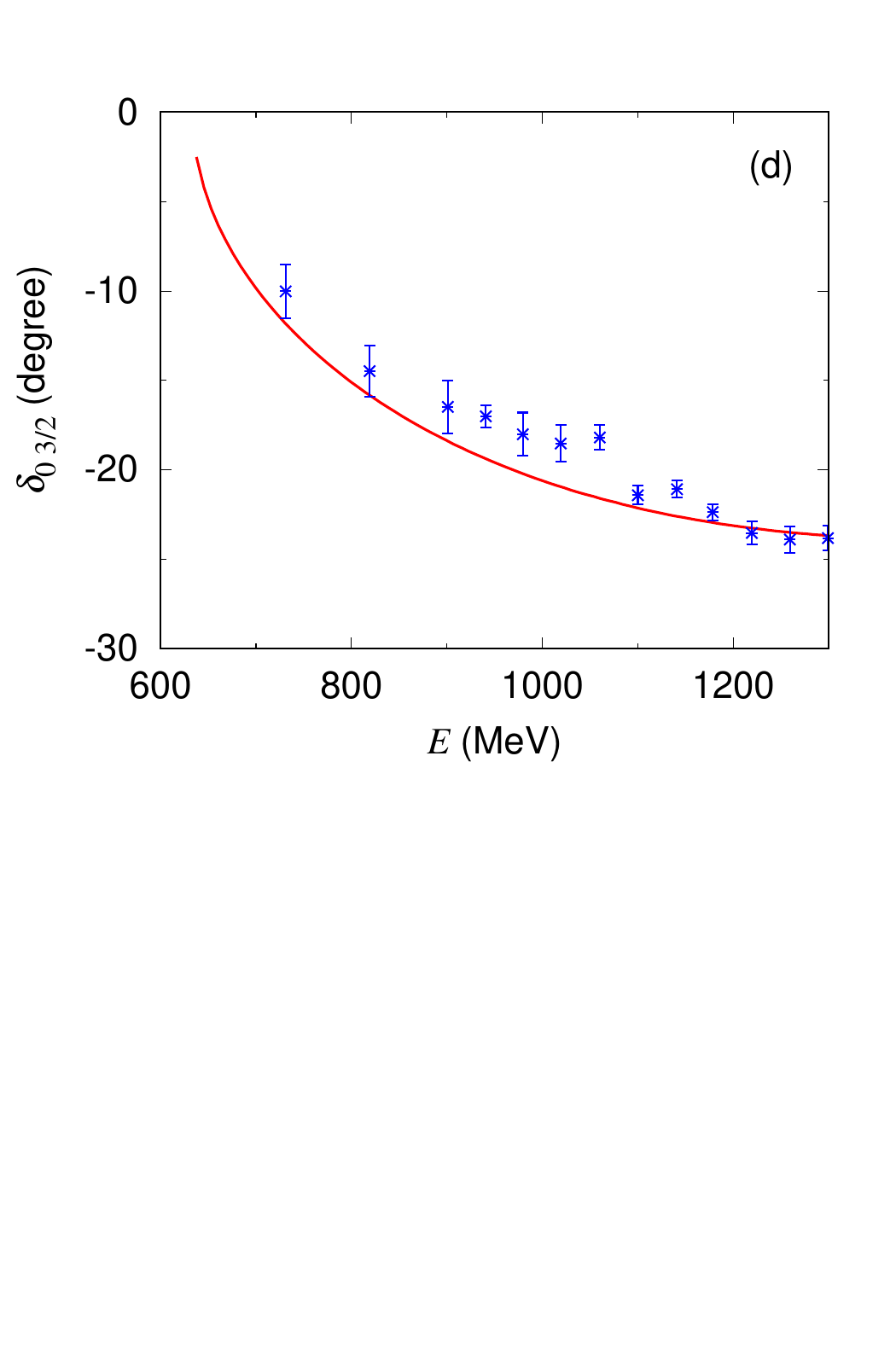}
\includegraphics[width=.32\textwidth]{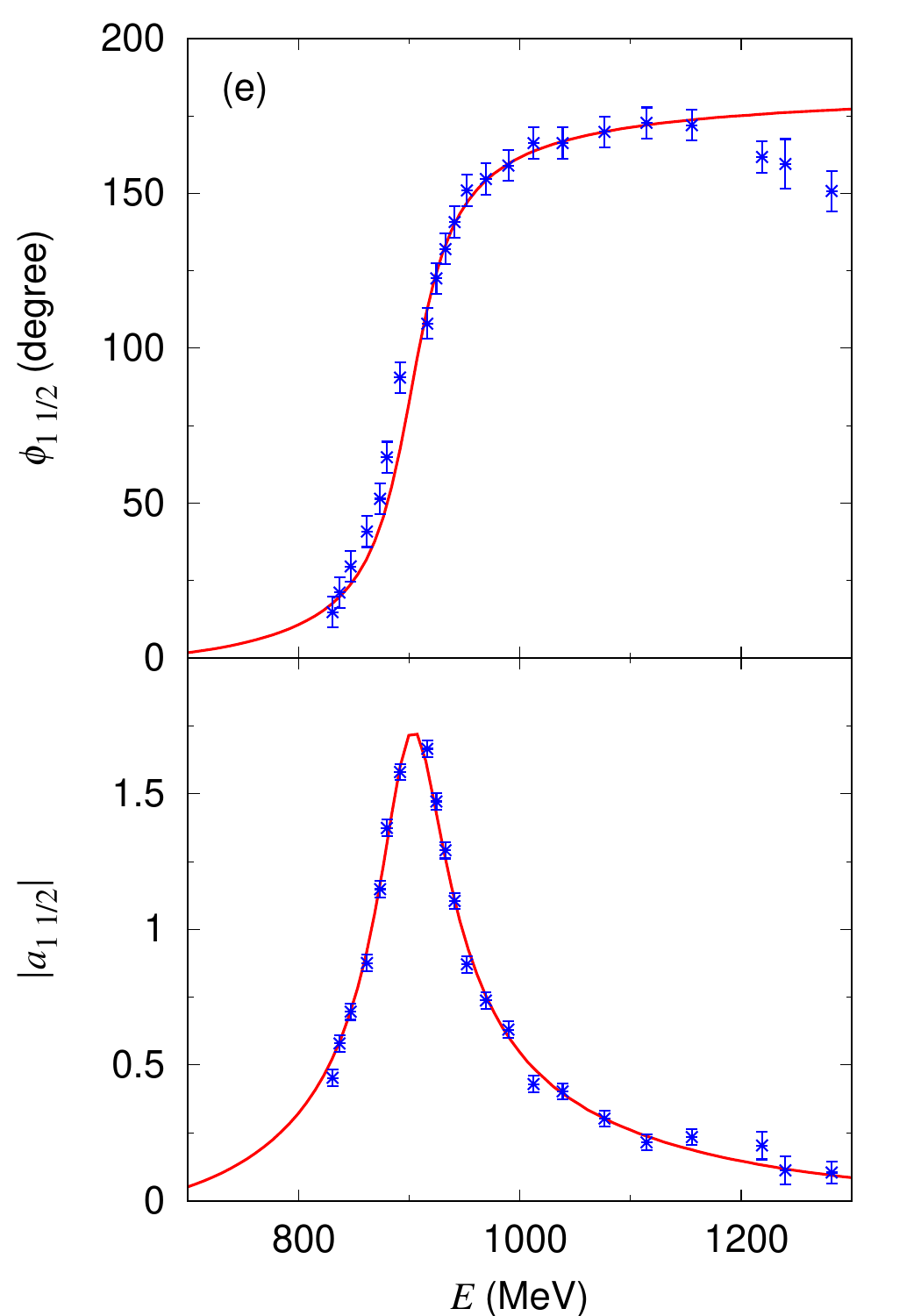}
\includegraphics[width=.32\textwidth]{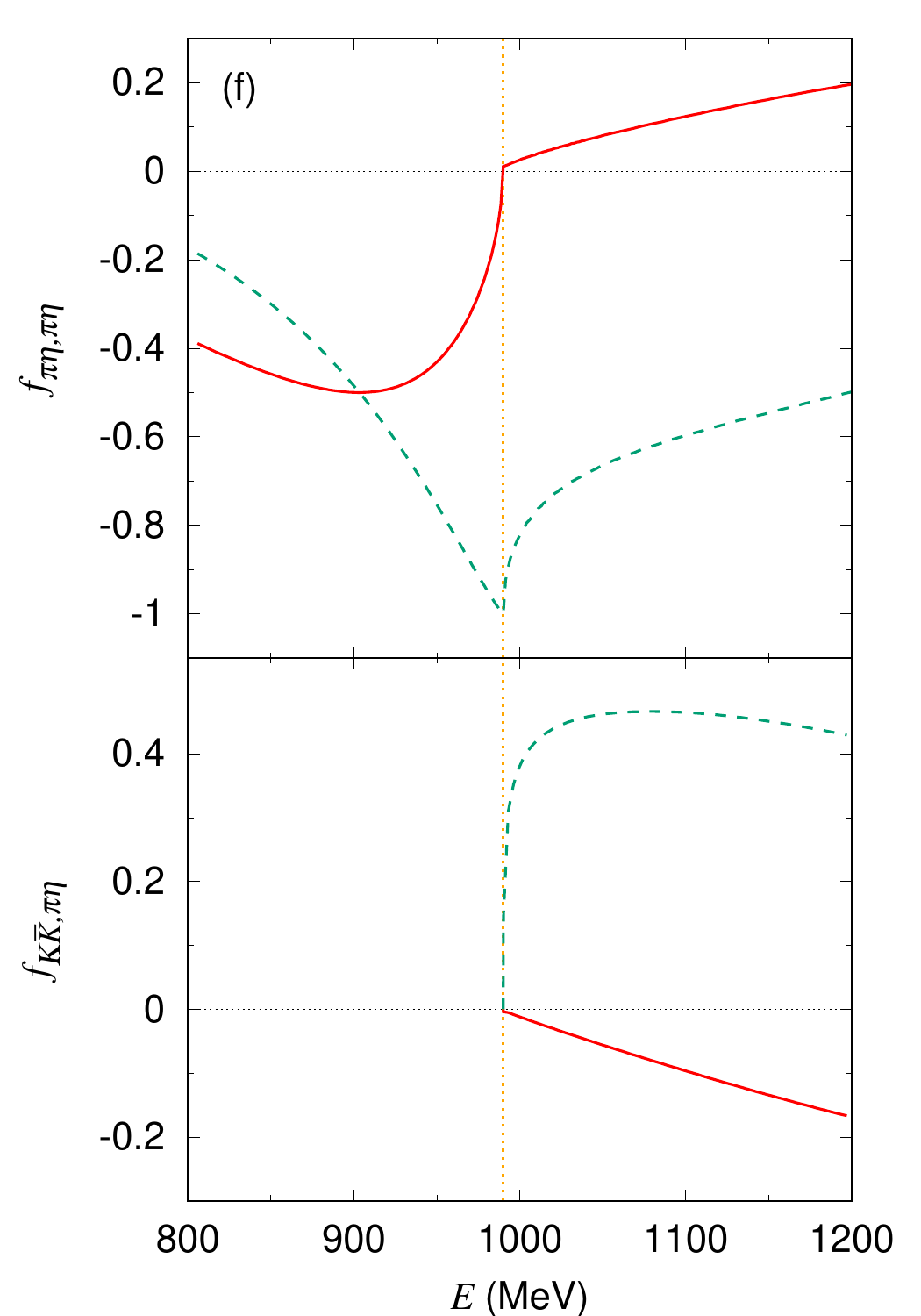}
\end{center}
 \caption{
(a,b) The $\pi\pi$ scattering. 
Phase shifts and inelasticities are shown in the upper and lower panels,
 respectively. Data are from Refs.~\cite{pipi-data1,pipi-data2,pipi-data3}.
(a) $\{L, I\}=\{0,0\}$; (b) $\{L, I\}=\{1,1\}$.
(c,e) The $\pi\bar{K}$ scattering.
Phase and modulus of the amplitudes are shown in the upper and lower panels,
 respectively. Data are from Ref.~\cite{LASS}.
(c) $L=0$ for $\pi^+K^-$; (e) $\{L, I\}=\{1,1/2\}$.
(d) Phase shifts of the $\pi K$ scattering
for $\{L, I\}=\{0,3/2\}$; data are from Ref.~\cite{pik-3hf}.
(f) The $\pi\eta\to \pi\eta$ (upper) and $\pi\eta\to K\bar{K}$ (lower)
scattering amplitudes; $\{L, I\}=\{0,1\}$. 
The real and imaginary parts are shown by the red solid and green dashed curves,
 respectively. 
The dotted vertical line indicates the $K\bar{K}$ threshold.
 }
\label{fig:two-body}
\end{figure*}

In our unitary coupled-channel model 
for describing the radiative $J/\psi$ decays in the $\eta(1405/1475)$
region,
$\pi\pi-K\bar{K}$, $\pi K$, and $\pi\eta-K\bar{K}$ coupled-channel scattering amplitudes 
of $E\ltap 1.2$~GeV are the major components. 
Our choices for the scattering models such as the number of
$R$ and contact interactions are specified in 
Table~\ref{tab:R}.
We determine the parameters in the two-meson scattering models
such as $h^{LI}_{a'b',ab}$, $b^{LI}_{ab}$, $m_R$, $g_{ab,R}$, and $c_{ab,R}$ 
in Eqs.~(\ref{eq:cont-ptl}), (\ref{eq:pw-2body-v}),
and (\ref{eq:pipi-vertex})
using experimental information.
For the $\pi\pi-K\bar{K}$ and $\pi K$ $s$- and $p$-wave scattering amplitudes,
we fit empirical scattering amplitudes by adjusting the model
parameters, and 
obtain reasonable fits as 
seen in Fig.~\ref{fig:two-body}(a-e).

Regarding the $\pi\eta-K\bar{K}$ $s$-wave scattering amplitude that includes the
$a_0(980)$ pole, we consider two experimental inputs.
First, our $a_0(980)$ propagator
($\tau^{LI}_{R',R}$ in Eq.~(\ref{eq:pw-2body-t}))
is fitted to 
the denominator of 
the $a_0(980)$ amplitude [Eq.~(4) of Ref.~\cite{bes3_chic1}]
from the BESIII amplitude analysis on $\chi_{c1}\to\eta\pi^+\pi^-$.
Second,
the ratio of coupling strengths (including the form factor) between the 
$a_0(980)\to \pi\eta$ and 
$a_0(980)\to K\bar{K}$ is fitted to an empirical value of 1.03 from Ref.~\cite{a0_980_ppbar}.
Furthermore, the relative phase between the $\pi\eta\to\pi\eta$ and 
$\pi\eta\to K\bar{K}$ amplitudes is chosen to be consistent with those
from the chiral unitary model~\cite{chUT}.
In Fig.~\ref{fig:two-body}(f),
we show 
our $\pi\eta\to\pi\eta$ and 
$\pi\eta\to K\bar{K}$ scattering amplitudes defined by
\begin{eqnarray}
f^{LI}_{a'b',ab} (E) &&= \pi
 \sqrt{\rho_{a'b'}\, {\cal B}_{a'b'}}
 \sqrt{\rho_{ab}\, {\cal B}_{ab}}
 T^{LI}_{a'b',ab} (q'_o,q_o; E) \ .
\nonumber\\
\label{eq:famp}
\end{eqnarray}

Finally, we obtain the $\pi\eta-K\bar{K}-\rho\pi$ $d$-wave scattering amplitude
with the $a_2(1320)$ pole by
adjusting the model parameters so that 
the mass and width of $a_2(1320)$, and branching fractions of
$a_2(1320)\to\pi\eta$ and 
$a_2(1320)\to K\bar{K}$ are reproduced;
all of the fitted $a_2(1320)$ properties are from 
the PDG listing~\cite{pdg}.

From the obtained partial wave amplitudes, resonance
poles are extracted and presented in 
Tables~\ref{tab:pole-pipi} -\ref{tab:pole-pieta}.
Overall, the pole locations are consistent with those listed in the PDG~\cite{pdg}.
Numerical values of the fitting parameters
are given in Tables~\ref{tab:kpi}--\ref{tab:pieta}.

\section{Parameters fitted to radiative $J/\psi$ decay data}
\label{app2}

Table~\ref{tab:param} presents model parameters determined by fitting 
$J/\psi\to\gamma (0^{-+})\to\gamma(K_SK_S\pi^0)$ Dalitz plot pseudodata
and the branching fractions of 
$\eta(1405/1475)\to\eta\pi^+\pi^-$ and 
$\eta(1405/1475)\to\rho^0\gamma$
relative to that of 
$\eta(1405/1475)\to K\bar{K}\pi$.
When a two-meson scattering model includes contact interactions, 
we consider a direct bare $M^*\to abc$ decay where 
two pseudoscalar mesons ($ab$) have an orbital angular momentum $L$
and a total isospin $I$.
We describe this bare vertex function with
[cf. Eq.~(\ref{eq:bare_mstar})]
\begin{eqnarray}
F_{(c(ab)_{LI})_{l},M^*_i}(q)\! &=& 
C^{M^*_i}_{(c (ab)_{LI})_l}
\left({q\over m_\pi}\right)^l
\nonumber\\
&&\times 
\frac{ [1+q^2/(\Lambda^{M^*_i}_{(c (ab)_{LI})_l})^2]^{-2-{l\over 2}} 
}
{\sqrt{4 E_c(q) m_{M^*_i}}} 
,
\label{eq:bare_mstar-contact}
\end{eqnarray}
where 
$C^{M^*_i}_{(c (ab)_{LI})_l}$ and 
$\Lambda^{M^*_i}_{(c (ab)_{LI})_l}$
are coupling and cutoff parameters, respectively. 
This bare vertex function is used in a dressed vertex and a self energy
in a similar manner as the bare vertex
$F_{(cR)_{l},M^*_i}$ in Eq.~(\ref{eq:bare_mstar})
is used in Eqs.~(\ref{eq:dressed_mstar}), (\ref{eq:dressed-g}), 
and (\ref{eq:mstar-sigma}).

\begin{table}
\renewcommand{\arraystretch}{1.6}
\tabcolsep=4.mm
\caption{\label{tab:pole-pipi} 
Pole positions ($M_{\rm pole}$) in our $\pi\pi$ scattering amplitudes. 
The Riemann sheets (RS) of the pole positions are specified by $(s_{\pi\pi},s_{K\bar{K}})$
where $s_{x}=p(u)$ indicates that a pole is on the physical
 (unphysical) sheet of the channel $x$; ``$-$'' (hyphen) indicates no coupling to
 the channel.
}    
\begin{tabular}{lccc}
$\{L,I\}$& $M_{\rm pole}$ (MeV)  & RS & name \\\hline
\multirow{3}{*}{\{0, 0\}} 
         & $438 - 311i$ & $(up)$ & $\sigma$ \\
         & $1000-  20i$ & $(up)$ & $f_0(980)$ \\
         & $1420 -224i$ & $(uu)$ & $f_0(1370)$ \\\hline
\{1, 1\} & $769  - 78i$ & $(u-)$  & $\rho(770)$ \\
\end{tabular}
\end{table}

\begin{table}
\renewcommand{\arraystretch}{1.6}
\tabcolsep=4.mm
\caption{\label{tab:pole-piK} 
Pole positions ($M_{\rm pole}$) in our $\pi K$ scattering amplitudes. 
The Riemann sheets (RS) of the pole positions are specified by $(s_{\pi K})$.
}    
\begin{tabular}{lccc}
$\{L,I\}$& $M_{\rm pole}$ (MeV)  & RS & name \\\hline
\{0, 1/2\} & $665-297i$ & $(u)$ & $\kappa$ \\
\{1, 1/2\} & $902-30i$  & $(u)$ & $K^*(892)$ \\
\end{tabular}
\end{table}

\begin{table}
\renewcommand{\arraystretch}{1.6}
\tabcolsep=4.mm
\caption{\label{tab:pole-pieta} 
Pole positions ($M_{\rm pole}$) in our $\pi\eta$ scattering amplitudes. 
The Riemann sheets (RS) of the pole positions are specified by
 $(s_{\pi\eta},s_{K\bar{K}},s_{\rho\pi})$.
}    
\begin{tabular}{lccc}
$\{L,I\}$& $M_{\rm pole}$ (MeV)  & RS & name \\\hline
\{0, 1\} & $1070-112i$ & $(up-)$  & $a_0(980)$ \\
\{2, 1\} & $1322-56i$  & $(uuu)$ & $a_2(1320)$ \\
\end{tabular}
\end{table}

\begin{table}
\caption{\label{tab:kpi} 
Parameter values for the $\pi K$ partial wave scattering models.
The $i$-th bare $R$ states ($R_i$) has a mass of $m_{R_i}$, 
and it decays into $h_1$ and $h_2$ particles 
with couplings ($g_{h_1h_2,R_i}$)
and cutoffs ($c_{h_1h_2,R_i}$).
Couplings and cutoffs for contact interactions
are denoted by $h_{h_1h_2,h_1h_2}$ and 
$b_{h_1h_2}$, respectively.
The parameters have been defined in 
Eqs.~(\ref{eq:cont-ptl}), (\ref{eq:vf-cont}), 
(\ref{eq:pw-2body-v}), and (\ref{eq:pipi-vertex}).
For simplicity, we suppress
the superscripts, $LI$, of the parameters.
The mass and cutoff values are given in the unit of MeV,
and the couplings are dimensionless.
}    
\begin{ruledtabular}
\begin{tabular}{lrrr}
$R~\{L,I\}$&$\kappa({K}^*_0)$ \{0, 1/2\} & \{0, 3/2\} &${K}^*$ \{1, 1/2\} \\\hline
$m_{R_1}$                  & 1239&      --- &    926  \\
$g_{\pi\bar{K},R_1}$       & 5.79&      --- &   0.74  \\ 	 
$c_{\pi\bar{K},R_1}$       & 1000&      --- &    752  \\ 
$h_{\pi\bar{K},\pi\bar{K}}$& 0.59&      0.47& $-$0.01 \\	 
$b_{\pi\bar{K}}$           & 1000&      1973&     752 \\
\end{tabular}
\end{ruledtabular}
\end{table}

\begin{table}
\caption{\label{tab:pipi} 
Parameter values for the $\pi\pi$ partial wave scattering models.
See Table~\ref{tab:kpi} for the description.
}
\begin{ruledtabular}
\begin{tabular}{lrr}
$R~\{L,I\}$&$f_0$ \{0, 0\}&$\rho$ \{1, 1\}  \\\hline
$m_{R_1}$          &    1007&         834 \\
$g_{\pi\pi,R_1}$   &    6.76&        1.03 \\
$c_{\pi\pi,R_1}$   &    1458&        1040 \\
$g_{K\bar{K},R_1}$ & $-$4.75&       --- \\ 
$c_{K\bar{K},R_1}$ &     711&         --- \\  
$m_{R_2}$ 	   &    1677&         --- \\
$g_{\pi\pi,R_2}$   & $-$5.87&       --- \\
$c_{\pi\pi,R_2}$   &    1458&         --- \\
$g_{K\bar{K},R_2}$ &   10.21&         --- \\ 
$c_{K\bar{K},R_2}$ &     711&         --- \\
$h_{\pi\pi,\pi\pi}$    &    0.65 &         --- \\	 
$h_{\pi\pi,K\bar{K}}$  & $-$0.42 &         --- \\	 
$h_{K\bar{K},K\bar{K}}$& $-$1.11 &         --- \\	 
$b_{\pi\pi}$           &    1458 &         --- \\
$b_{K\bar{K}}$         &     711 &         --- \\
\end{tabular}
\end{ruledtabular}
\end{table}

\begin{table}
\caption{\label{tab:pieta} 
Parameter values for the $\pi\eta$ partial wave scattering models.
See Table~\ref{tab:kpi} for the description.
}
\begin{ruledtabular}
\begin{tabular}{lrr}
$R~\{L,I\}$&$a_0$ \{0, 1\}&$a_2$ \{2, 1\}  \\\hline
$m_{R_1}$          &    1233 &  1436  \\
$g_{\pi\eta,R_1}$  & $-$3.08 &  0.09  \\
$c_{\pi\eta,R_1}$  &    1973 &  1000  \\
$g_{K\bar{K},R_1}$ &    2.94 &  0.07  \\ 
$c_{K\bar{K},R_1}$ &    1973 &  1000  \\  
$g_{\rho\pi,R_1}$  & ---     &  0.33  \\ 
$c_{\rho\pi,R_1}$  & ---     &  1000  \\  
\end{tabular}
\end{ruledtabular}
\end{table}

\begin{table*}[t]
\vspace*{-3mm}
\caption{\label{tab:param}
Parameter values for $i$-th bare $\eta^*$ state
obtained from one of the bootstrap fits.
The symbols are
the $J/\psi\eta^*_i\gamma$ coupling constant 
$g_{J/\psi\eta^*_i\gamma}$ in Eq.~(\ref{eq:invamp2}),
the bare mass $m_{\eta^*_i}$ in Eq.~(\ref{eq:mstar-g1}),
and bare couplings 
$C^i_{c {R}^n}$ in Eq.~(\ref{eq:bare_mstar})
and $C^i_{c (ab)_{LI}}$ in Eq.~(\ref{eq:bare_mstar-contact});
 the subscripts $l$ are suppressed.
$R^{n}$ stands for $n$-th bare $R$ state, while 
$(ab)_{LI}$ is a direct decay into two pseudoscalar
mesons ($ab$) with the orbital angular momentum $L$
and total isospin $I$.
The nonresonant amplitude parameter,
$c_{\rm NR}$, has been introduced in Eq.~(\ref{eq:cnr}).
Since the overall normalization of the full amplitude is arbitrary in
 our model, a common scaling factor can be multiplied to
$g_{J/\psi\eta^*_i\gamma}$ and $c_{\rm NR}$.
All cutoffs 
[$\Lambda^i_{c {R}^n}$ in Eq.~(\ref{eq:bare_mstar})
and $\Lambda^i_{c (ab)_{LI}}$ in Eq.~(\ref{eq:bare_mstar-contact})]
are fixed to 700~MeV.
}
\renewcommand{\arraystretch}{1.3}
\begin{ruledtabular}
\begin{tabular}{lrlr}
   $m_{\eta^*_1}$ (MeV)              &$ 1622   $&   $m_{\eta^*_2}$  (MeV)          &$ 2309   $\\
   $g_{J/\psi\eta^*_1\gamma}$        &1 (fixed) &   $g_{J/\psi\eta^*_2\gamma}$   &$  0.317     -0.454    i$\\
   $C^1_{\bar{K}K^*}$                &$   2.00    $ &$C^2_{\bar{K}K^*}$                &$   4.54    $\\
   $C^1_{\bar{K}(\pi K)_{1{1\over 2}}}$       &$ -0.421    $ &$C^2_{\bar{K}(\pi K)_{1{1\over 2}}}$       &$ -0.124    $\\
   $C^1_{\bar{K}\kappa}$             &$   19.4    $ &$C^2_{\bar{K}\kappa}$             &$   22.6    $\\
   $C^1_{\bar{K}(\pi K)_{0{1\over 2}}}$       &$   3.10    $ &$C^2_{\bar{K}(\pi K)_{0{1\over 2}}}$       &$   2.19    $\\
   $C^1_{\pi a_0}$                   &$  -1.01    $ &$C^2_{\pi a_0}$                   &$  -1.15    $\\
   $C^1_{\pi a_2}$                   &$   2.38    $ &$C^2_{\pi a_2}$                   &$  -3.12    $\\
   $C^1_{\eta f^1_0}$                &$  -2.40    $ &$C^2_{\eta f^1_0}$                &$  -7.28    $\\
   $C^1_{\eta f^2_0}$                &$  -1.33    $ &$C^2_{\eta f^2_0}$                &$   9.26    $\\
   $C^1_{\eta(\pi\pi)_{00}}$         &$  0.358    $ &$C^2_{\eta(\pi\pi)_{00}}$         &$   3.10    $\\
   $C^1_{\eta(K\bar{K})_{00}}$       &0 (fixed) &$C^2_{\eta(K\bar{K})_{00}}$       &0 (fixed) \\
   $C^1_{\rho\rho}$                  &0 (fixed) &$C^2_{\rho\rho}$                  &$   44.9    $\\
   $c_{\rm NR}$ (GeV$^{-2}$)         &  $  127 - 41 i$&&\\
\end{tabular}
\end{ruledtabular}
\end{table*}

\clearpage



\end{document}